\documentclass[12pt]{article}
\usepackage{graphicx,amssymb,amsmath,empheq}
\usepackage{authblk}
\usepackage{amsthm,bm}
\usepackage{empheq}
\usepackage{subfig} 

\usepackage{pifont}
\def\X#1{%
        \ding{\numexpr171+#1\relax}%
}

\usepackage{hyperref}
\hypersetup{colorlinks = true, linkcolor=black, citecolor=black, urlcolor=blue}
\usepackage{url}

\theoremstyle{plain}

\usepackage{tikz}
\usepackage{tikz-cd}
\usetikzlibrary{arrows}
\usetikzlibrary{intersections}
\usetikzlibrary{shapes.geometric}
\usetikzlibrary{decorations.pathmorphing, patterns,shapes}
\usetikzlibrary{decorations.markings}

\tikzset{
  mid arrow/.style={postaction={decorate,decoration={
        markings,
        mark=at position .575 with {\arrow[#1]{stealth}}
      }}},
  near arrow/.style={postaction={decorate,decoration={
        markings,
        mark=at position .375 with {\arrow[#1]{stealth}}
      }}},
   far arrow/.style={postaction={decorate,decoration={
        markings,
        mark=at position .7 with {\arrow[#1]{stealth}}
      }}},
}

\tikzset{
  baseline = -0.5ex,
  wavy/.style = {
    thick,
    decorate,
    decoration={snake,amplitude=2pt,segment length=5pt}},
  sdot/.style = {
    circle,
    draw=none,
    fill=black,
    minimum size=2.5pt,
    inner sep=0pt},
  bdot/.style = {
    circle,
    draw=none,
    fill=black,
    minimum size=4pt,
    inner sep=0pt},
  svertex/.style = {
    circle,
    draw=black,
    thick,
    fill=lightgray,
    minimum size=8pt,
    inner sep=1pt},
  mvertex/.style = {
    circle,
    draw=black,
    thick,
    fill=lightgray,
    minimum size=12pt,
    inner sep=1pt},
  bvertex/.style = {
    circle,
    draw=black,
    thick,
    fill=lightgray,
    minimum size=16pt}}

\usepackage[nosort]{cite}

\renewcommand{\bar}{\overline}
\renewcommand{\tilde}{\widetilde}

\renewcommand{\geq}{\geqslant}

\newcommand{\dkap}{\delta\kern-1.25pt\varkappa}

\makeatletter

\newcommand*{\wideboxed}[1]{\setlength{\fboxsep}{1ex}%
  \fbox{\m@th$\displaystyle#1$}}
\makeatother

\title{
Perturbative Page Curve Induced by External Impulse
}

\author[1]{Pengfei Zhang}
\affil[1]{\normalsize \it Department of Physics, Fudan University, Shanghai, 200438, China}

\date{\today}

\begin{document}
  \maketitle
  
  \begin{abstract}
In this work, we extend the recent study of entropy dynamics induced by an external impulse in open quantum systems, where the entropy response follows the Page curve. For small system-bath coupling $\kappa$, we expect that the entropy first increases exponentially $\kappa^2 e^{\varkappa t}$ in the early-time regime $t\lesssim |\log \kappa|$ due to quantum many-body chaos, and then decreases as $~e^{-\lambda_0 t}$ with $\lambda_0 \propto \kappa^2$ due to the energy relaxation. These results are confirmed through explicit calculations using two methods: 1). generalized Boltzmann equation for systems with quasi-particles; 2). scramblon effective theory in the early-time regime and perturbation theory in the late-time regime for 0+1-d systems. We also prove that in the second stage, the entropy of the system is equal to the coarse-grained entropy.

  \end{abstract}
  \tableofcontents

\section{Introduction}
Understanding the entropy of quantum many-body systems is a crucial topic in both high energy and condensed matter physics. Recent studies have highlighted the significance of the Page curve \cite{Page:1993df} in entropy dynamics. The Page curve describes the entanglement entropy of subsystem $A$, averaged over random pure states. Mathematically, the result can be expressed as $S_A^{\text{Page}}=\text{min}\{L_A,L-L_A\} \ln d+O(d^{-|2L_A-L|})$, where $L_A$ represents the subsystem size, $L$ denotes the total system size, and $d$ represents the local Hilbert space dimension. In high energy physics, the Page curve is connected to the evaporation process of pure-state black holes. As time progresses, the number of qubits in the black hole diminishes, and time $t$ can be seen as analogous to the subsystem size $L_A$. Consequently, one would expect the entropy of the black hole to initially increase and then decrease. However, Hawking's calculation yields a result that exhibits monotonic entropy growth. This discrepancy is known as the information paradox \cite{Hawking:1975vcx}. Recently, the information paradox has been resolved through the quantum generalization of the Ryu-Takayanagi formula \cite{Ryu:2006bv,Ryu:2006ef,Lewkowycz:2013nqa,Hubeny:2007xt,Faulkner:2013ana,Engelhardt:2014gca}, incorporating the contributions of islands due to replica wormholes \cite{almheiri2020replica,Penington:2019kki,almheiri2019islands}.

On the other hand, recent studies in condensed matter systems have focused on the entropy dynamics of initial states with short-range entanglement at infinite temperature \cite{Nahum:2016muy,PhysRevX.8.021013,PhysRevX.8.031057,PhysRevX.8.031058}. Under unitary evolutions generated by chaotic Hamiltonians \cite{PhysRevB.98.014309}, the steady state can be approximated as a random pure state, exhibiting a volume-law entanglement entropy in accordance with the Page curve. For states with finite energy density, quantum thermalization predicts that the slope of the curve should be replaced with the thermal entropy density \cite{PhysRevA.43.2046,PhysRevE.50.888}. Subsequent studies have revealed that repeated measurements can induce transitions in the entanglement entropy of the steady state \cite{Li_2018,Cao_Tilloy_2019,Li_2019,Skinner_2019,Chan_2019,gullans2019dynamical,gullans2019scalable,zabalo2019critical,Szyniszewski_2019}. Unlike in holographic systems, where entropy can be geometrically understood, most of these works rely on numerical simulations of circuit models, while the available analytical techniques remain quite limited. This limitation is partly attributed to the nonlinear nature of both the Von Neumann and the R\'enyi entropies. 

Response theory is an important route to uncover the underlying physics in quantum many-body systems. This includes both the conventional linear response theory \cite{altland2010condensed} and the more recent development of non-Hermitian linear response for open quantum systems \cite{pan2020non}. Consequently, there is a growing interest in developing entropy response theory for general quantum systems, with the potential for analytical solutions. Pioneering works in this direction include \cite{dadras2021perturbative,Chen:2020atj}. In particular, in \cite{dadras2021perturbative}, the authors investigate the impact of an external impulse on the entropy of open systems prepared in the thermal ensemble. This initial perturbation gives rise to excitations that are progressively absorbed by the bath, analogous to black hole evaporation. As a result, the entropy response exhibits characteristics similar to the Page curve, leading to the notion of a ``perturbative Page curve''. In their study, the authors compute the early-time exponential growth of the von Neumann entropy by establishing a connection with ``branched'' out-of-time-order correlators. In this work, we further extend the study beyond the regime of exponential growth by employing a general non-Markovian model of the bath.

The paper is organized as follows: In Section \ref{secII}, we begin by describing the setup of our problem, where we apply an external impulse to open quantum systems in thermal equilibrium. We then elucidate the two stages of entropy dynamics by establishing their connection to two-point functions on the entropy contour.
In Section \ref{secIII}, we develop an approach based on the Boltzmann equation to compute the corresponding two-point functions. This method is applicable to systems with quasi-particles. Additionally, we present numerical results for a specific example in the context of 0+1-dimensional systems.
Subsequently, in Section \ref{secIV}, we extend our study to more general 0+1-dimensional systems, which may describe non-Fermi liquids lacking quasi-particles.
Finally, in Section \ref{secFinal}, we conclude the paper with discussions on future directions for research.

\section{Entropy After an External Impulse}\label{secII}
In this section, we describe the protocol considered in this paper, which contains applying an external impulse to open quantum systems in thermal equilibrium. We relate the presence of a perturbative Page curve to two-point functions on the entropy contour, which is dominated by quantum many-body chaos in the early-time regime, and by energy relaxation in the late-time regime. 

\subsection{The set-up}
We first describe the set-up proposed in \cite{dadras2021perturbative} which exhibits the perturbative Page curve. We consider a quantum system described by some Hamiltonian $H_s$. The system is coupled to a large external heat bath $b$ with Hamiltonian $H_b$ through a coupling term $H_{sb}$:
\begin{equation}
H_{sb}=\kappa\sum_{i=1}^NO_s^iO_b^i.
\end{equation} 
Here $O_s^i/O_b^i$ is an operator on system/bath. The total Hamiltonian $H$ reads $H=H_s+H_b+H_{sb}$. We consider preparing the total system in a thermal ensemble at finite inverse temperature $\beta$, decribed by the density matrix $\rho_0=Z^{-1}e^{-\beta H}$ with $Z=\text{tr}~e^{-\beta H}$. This can be equivalently understood as the late-time phase of an initial thermofield double (TFD) state, in which the entropy of system $s$ saturates \cite{dadras2021perturbative}. 

We are interested in the effect of an external impulse on the entropy $S$ of system $s$. We model the impulse by introducing a perturbation
\begin{equation}
\rho(t=0,\epsilon)=e^{-i\epsilon X}\rho_0 e^{i\epsilon X},
\end{equation}
Here $X$ is a Hermitian operator acting on the system $s$. The total system then evolves as 
\begin{equation}
\rho(t,\epsilon)=e^{-iHt}e^{-i\epsilon X}\rho_0 e^{i\epsilon X}e^{iHt}.
\end{equation}
The $n$-th R\'enyi entropy of system $s$ can be computed as
\begin{equation}
S^{(n)}(t,\epsilon)=\frac{1}{1-n}\log\big[\text{tr}_s~\rho_s(t,\epsilon)^n\big],\ \ \ \ \ \  \rho_s(t,\epsilon)=\text{tr}_b~\rho(t,\epsilon).
\end{equation}
In particular, the Von Neumann entropy $S(t,\epsilon)$ is determined by taking the limit of $n\rightarrow 1$.

Without the perturbation, the entropy $S^{(n)}(t,0)=S^{(n)}_0$ is the thermal subsystem entropy, which includes both the thermal entropy of system $s$, and the entanglement entropy between $s$ and $b$. We are interested in the effect of small perturbation $\epsilon \ll1$ on the entropy $S^{(n)}(t,\epsilon)$. Similar to \cite{dadras2021perturbative}, we focus on the second order derivative 
\begin{equation}
\delta S^{(n)}(t)=\left.\frac{1}{2}\frac{\partial^2}{\partial \epsilon^2}S^{(n)}(t,\epsilon)\right|_{\epsilon=0}.
\end{equation}

\subsection{Physical expectation and Green's functions}
As described in \cite{dadras2021perturbative}, the evolution of $\delta S^{(n)}(t)$ contains two stages. After applying the unitary operator $e^{-i\epsilon X}$, the entropy begins to increase smoothly since the perturbation creates excitations that entangle the system and bath through the system-bath coupling $H_{sb}$. Studies relate that such entropy dynamics to OTOCs \cite{Fan:2016ean,Hosur:2015ylk}, which leads to an exponential growth $\sim \kappa^2 e^{\varkappa t}$ at early time \cite{dadras2021perturbative}. Here $\varkappa$ is the quantum Lyapunov exponent. The saturation occurs when $\varkappa t\sim -\log \kappa$. Afterwards, the dynamics is dominate by the relaxation of the energy, and the entropy is equal to the coarse-grained entropy. In the long-time limit, the total system reaches the thermal equilibrium again at inverse temperature $\beta$. As a result, $S^{(n)}(\infty,x)=S^{(n)}(0,x)=S^{(n)}_0$. 

In this subsection, we hope to understand how these two stages appear mathematically for $\delta S^{(n)}(t)$. For conciseness, we focus on the second R\'enyi entropy, which corresponds to $n=2$. The generalization to arbitrary $n$ is straightforward. We first consider utilizing the path-integral approach for the unperturbed entropy $S^{(2)}(t,0)$. A pictorial representation reads 
\begin{equation}\label{eqn:entropycontour}\exp\left(-S^{(2)}(t,0)\right)=\text{tr}_s\big[\text{tr}_b~e^{-iHt}\rho_0 e^{iHt}\big ]^2=
\begin{tikzpicture}[scale = 0.75,baseline={([yshift=-3.2pt]current bounding box.center)}]
   \draw[blue,thick] (-0.6,0) arc(180:8:0.6 and 0.6);
   \draw[blue,thick] (-0.6,0) arc(180:352:0.6 and 0.6);
   \draw[black,thick] (-0.8,0) arc(180:15:0.8 and 0.8);
   \draw[black,thick] (-0.8,0) arc(180:345:0.8 and 0.8);
      \draw[blue] (-0.25,0.25) node{\scriptsize$b$};
      \draw[black] (-0.725,0.725) node{\scriptsize$s$};
   \draw[black,thick,mid arrow] (0.77,0.2)-- (2,0.2);
   \draw[black,thick,mid arrow] (2,-0.2)--(0.77,-0.2);
   \draw[blue,thick] (0.58,0.075)-- (1.9,0.075);
   \draw[blue,thick] (1.9,-0.075)-- (0.58,-0.075);  
      
   \draw[blue,thick] (1.9,0.075) arc(90:-90:0.075 and 0.075);

      \draw[blue,thick] (-0.6,-1.8) arc(180:8:0.6 and 0.6);
   \draw[blue,thick] (-0.6,-1.8) arc(180:352:0.6 and 0.6);
   \draw[black,thick] (-0.8,-1.8) arc(180:15:0.8 and 0.8);
   \draw[black,thick] (-0.8,-1.8) arc(180:345:0.8 and 0.8);
      \draw[blue] (-0.25,-1.55) node{\scriptsize$b$};
      \draw[black] (-0.725,-1.075) node{\scriptsize$s$};
   \draw[black,thick,mid arrow] (0.77,-1.6)-- (2,-1.6);
   \draw[black,thick,mid arrow] (2,-2)--(0.77,-2);
   \draw[blue,thick] (0.58,-1.725)-- (1.9,-1.725);
   \draw[blue,thick] (1.9,-1.875)-- (0.58,-1.875);  

   \draw[black,thick] (1.98,-0.2)-- (1.98,-1.6);
    \draw[black,thick] (1.98,0.2)-- (1.98,0.6);
    \draw[black,thick] (1.98,-2)-- (1.98,-2.4);

    \draw[black,thick] (1.92,0.4)-- (2.04,0.4);
    \draw[black,thick] (1.92,-2.2)-- (2.04,-2.2);
      
   \draw[blue,thick] (1.9,-1.725) arc(90:-90:0.075 and 0.075);

   \filldraw[red] (0.8,0.2) circle (1.2pt) node {$ $};
    \filldraw[red] (0.8,-0.2) circle (1.2pt) node {$ $};

       \filldraw[red] (0.8,-1.6) circle (1.2pt) node {$ $};
      \filldraw[red] (0.8,-2) circle (1.2pt) node {$ $};

   \filldraw (0.95,0.2) circle (0pt) node[above] {\scriptsize$0$};
   \filldraw (1.75,0.2) circle (0pt) node[above] {\scriptsize$t$};
\end{tikzpicture}
\end{equation}
Here we use black/blue line for the contour of system $s$/bath $b$, which are coupled by the system-bath interaction $H_{sb}$. The red points represent the position where operators should be  inserted if we turn on finite $\epsilon$. Expanding $e^{-i\epsilon X}=(1-i\epsilon X-\epsilon^2\frac{X^2}{2}+O(\epsilon^3))$, we find six contributions:
\begin{equation}\label{eqn:sixcontributions}
\begin{aligned}
\delta S^{(2)}(t)&=\begin{tikzpicture}[scale = 0.75,baseline={([yshift=-3.2pt]current bounding box.center)}]
   \draw[blue,thick] (-0.6,0) arc(180:8:0.6 and 0.6);
   \draw[blue,thick] (-0.6,0) arc(180:352:0.6 and 0.6);
   \draw[black,thick] (-0.8,0) arc(180:15:0.8 and 0.8);
   \draw[black,thick] (-0.8,0) arc(180:345:0.8 and 0.8);
      \draw[blue] (-0.25,0.25) node{\scriptsize$b$};
      \draw[black] (-0.725,0.725) node{\scriptsize$s$};
   \draw[black,thick,mid arrow] (0.77,0.2)-- (2,0.2);
   \draw[black,thick,mid arrow] (2,-0.2)--(0.77,-0.2);
   \draw[blue,thick] (0.58,0.075)-- (1.9,0.075);
   \draw[blue,thick] (1.9,-0.075)-- (0.58,-0.075);  
      
   \draw[blue,thick] (1.9,0.075) arc(90:-90:0.075 and 0.075);

      \draw[blue,thick] (-0.6,-1.8) arc(180:8:0.6 and 0.6);
   \draw[blue,thick] (-0.6,-1.8) arc(180:352:0.6 and 0.6);
   \draw[black,thick] (-0.8,-1.8) arc(180:15:0.8 and 0.8);
   \draw[black,thick] (-0.8,-1.8) arc(180:345:0.8 and 0.8);
      \draw[blue] (-0.25,-1.55) node{\scriptsize$b$};
      \draw[black] (-0.725,-1.075) node{\scriptsize$s$};
   \draw[black,thick,mid arrow] (0.77,-1.6)-- (2,-1.6);
   \draw[black,thick,mid arrow] (2,-2)--(0.77,-2);
   \draw[blue,thick] (0.58,-1.725)-- (1.9,-1.725);
   \draw[blue,thick] (1.9,-1.875)-- (0.58,-1.875);  

   \draw[black,thick] (1.98,-0.2)-- (1.98,-1.6);
    \draw[black,thick] (1.98,0.2)-- (1.98,0.6);
    \draw[black,thick] (1.98,-2)-- (1.98,-2.4);

    \draw[black,thick] (1.92,0.4)-- (2.04,0.4);
    \draw[black,thick] (1.92,-2.2)-- (2.04,-2.2);
      
   \draw[blue,thick] (1.9,-1.725) arc(90:-90:0.075 and 0.075);

   \filldraw[red] (0.8,0.2) circle (1.2pt) node {$ $};
    \filldraw[red] (0.8,-0.2) circle (1.2pt) node {$ $};

       \filldraw[red] (0.8,-1.6) circle (1.2pt) node {$ $};
      \filldraw[red] (0.8,-2) circle (1.2pt) node {$ $};
     \draw[brown] (1.3,-0.9) node{\X1};

   \filldraw (1,0.2) circle (0pt) node[above] {\scriptsize$X^2$};
\end{tikzpicture}\times \frac{1}{2}\times 2+
\begin{tikzpicture}[scale = 0.75,baseline={([yshift=-3.2pt]current bounding box.center)}]
   \draw[blue,thick] (-0.6,0) arc(180:8:0.6 and 0.6);
   \draw[blue,thick] (-0.6,0) arc(180:352:0.6 and 0.6);
   \draw[black,thick] (-0.8,0) arc(180:15:0.8 and 0.8);
   \draw[black,thick] (-0.8,0) arc(180:345:0.8 and 0.8);
      \draw[blue] (-0.25,0.25) node{\scriptsize$b$};
      \draw[black] (-0.725,0.725) node{\scriptsize$s$};
   \draw[black,thick,mid arrow] (0.77,0.2)-- (2,0.2);
   \draw[black,thick,mid arrow] (2,-0.2)--(0.77,-0.2);
   \draw[blue,thick] (0.58,0.075)-- (1.9,0.075);
   \draw[blue,thick] (1.9,-0.075)-- (0.58,-0.075);  
      
   \draw[blue,thick] (1.9,0.075) arc(90:-90:0.075 and 0.075);

      \draw[blue,thick] (-0.6,-1.8) arc(180:8:0.6 and 0.6);
   \draw[blue,thick] (-0.6,-1.8) arc(180:352:0.6 and 0.6);
   \draw[black,thick] (-0.8,-1.8) arc(180:15:0.8 and 0.8);
   \draw[black,thick] (-0.8,-1.8) arc(180:345:0.8 and 0.8);
      \draw[blue] (-0.25,-1.55) node{\scriptsize$b$};
      \draw[black] (-0.725,-1.075) node{\scriptsize$s$};
   \draw[black,thick,mid arrow] (0.77,-1.6)-- (2,-1.6);
   \draw[black,thick,mid arrow] (2,-2)--(0.77,-2);
   \draw[blue,thick] (0.58,-1.725)-- (1.9,-1.725);
   \draw[blue,thick] (1.9,-1.875)-- (0.58,-1.875);  

   \draw[black,thick] (1.98,-0.2)-- (1.98,-1.6);
    \draw[black,thick] (1.98,0.2)-- (1.98,0.6);
    \draw[black,thick] (1.98,-2)-- (1.98,-2.4);

    \draw[black,thick] (1.92,0.4)-- (2.04,0.4);
    \draw[black,thick] (1.92,-2.2)-- (2.04,-2.2);
      
   \draw[blue,thick] (1.9,-1.725) arc(90:-90:0.075 and 0.075);

   \filldraw[red] (0.8,0.2) circle (1.2pt) node {$ $};
    \filldraw[red] (0.8,-0.2) circle (1.2pt) node {$ $};

       \filldraw[red] (0.8,-1.6) circle (1.2pt) node {$ $};
      \filldraw[red] (0.8,-2) circle (1.2pt) node {$ $};

     \draw[brown] (1.3,-0.9) node{\X2};
   \filldraw (1,-0.2) circle (0pt) node[below] {\scriptsize$X^2$};
\end{tikzpicture}\times \frac{1}{2}\times 2+\begin{tikzpicture}[scale = 0.75,baseline={([yshift=-3.2pt]current bounding box.center)}]
   \draw[blue,thick] (-0.6,0) arc(180:8:0.6 and 0.6);
   \draw[blue,thick] (-0.6,0) arc(180:352:0.6 and 0.6);
   \draw[black,thick] (-0.8,0) arc(180:15:0.8 and 0.8);
   \draw[black,thick] (-0.8,0) arc(180:345:0.8 and 0.8);
      \draw[blue] (-0.25,0.25) node{\scriptsize$b$};
      \draw[black] (-0.725,0.725) node{\scriptsize$s$};
   \draw[black,thick,mid arrow] (0.77,0.2)-- (2,0.2);
   \draw[black,thick,mid arrow] (2,-0.2)--(0.77,-0.2);
   \draw[blue,thick] (0.58,0.075)-- (1.9,0.075);
   \draw[blue,thick] (1.9,-0.075)-- (0.58,-0.075);  
      
   \draw[blue,thick] (1.9,0.075) arc(90:-90:0.075 and 0.075);

      \draw[blue,thick] (-0.6,-1.8) arc(180:8:0.6 and 0.6);
   \draw[blue,thick] (-0.6,-1.8) arc(180:352:0.6 and 0.6);
   \draw[black,thick] (-0.8,-1.8) arc(180:15:0.8 and 0.8);
   \draw[black,thick] (-0.8,-1.8) arc(180:345:0.8 and 0.8);
      \draw[blue] (-0.25,-1.55) node{\scriptsize$b$};
      \draw[black] (-0.725,-1.075) node{\scriptsize$s$};
   \draw[black,thick,mid arrow] (0.77,-1.6)-- (2,-1.6);
   \draw[black,thick,mid arrow] (2,-2)--(0.77,-2);
   \draw[blue,thick] (0.58,-1.725)-- (1.9,-1.725);
   \draw[blue,thick] (1.9,-1.875)-- (0.58,-1.875);  

   \draw[black,thick] (1.98,-0.2)-- (1.98,-1.6);
    \draw[black,thick] (1.98,0.2)-- (1.98,0.6);
    \draw[black,thick] (1.98,-2)-- (1.98,-2.4);

    \draw[black,thick] (1.92,0.4)-- (2.04,0.4);
    \draw[black,thick] (1.92,-2.2)-- (2.04,-2.2);
      
   \draw[blue,thick] (1.9,-1.725) arc(90:-90:0.075 and 0.075);

   \filldraw[red] (0.8,0.2) circle (1.2pt) node {$ $};
    \filldraw[red] (0.8,-0.2) circle (1.2pt) node {$ $};

       \filldraw[red] (0.8,-1.6) circle (1.2pt) node {$ $};
      \filldraw[red] (0.8,-2) circle (1.2pt) node {$ $};
           \draw[brown] (1.3,-0.9) node{\X3};

   \filldraw (1,0.2) circle (0pt) node[above] {\scriptsize$X$};
   \filldraw (1,-0.2) circle (0pt) node[below] {\scriptsize$X$};
\end{tikzpicture}\times (-2)\\
&+\begin{tikzpicture}[scale = 0.75,baseline={([yshift=-3.2pt]current bounding box.center)}]
   \draw[blue,thick] (-0.6,0) arc(180:8:0.6 and 0.6);
   \draw[blue,thick] (-0.6,0) arc(180:352:0.6 and 0.6);
   \draw[black,thick] (-0.8,0) arc(180:15:0.8 and 0.8);
   \draw[black,thick] (-0.8,0) arc(180:345:0.8 and 0.8);
      \draw[blue] (-0.25,0.25) node{\scriptsize$b$};
      \draw[black] (-0.725,0.725) node{\scriptsize$s$};
   \draw[black,thick,mid arrow] (0.77,0.2)-- (2,0.2);
   \draw[black,thick,mid arrow] (2,-0.2)--(0.77,-0.2);
   \draw[blue,thick] (0.58,0.075)-- (1.9,0.075);
   \draw[blue,thick] (1.9,-0.075)-- (0.58,-0.075);  
      
   \draw[blue,thick] (1.9,0.075) arc(90:-90:0.075 and 0.075);

      \draw[blue,thick] (-0.6,-1.8) arc(180:8:0.6 and 0.6);
   \draw[blue,thick] (-0.6,-1.8) arc(180:352:0.6 and 0.6);
   \draw[black,thick] (-0.8,-1.8) arc(180:15:0.8 and 0.8);
   \draw[black,thick] (-0.8,-1.8) arc(180:345:0.8 and 0.8);
      \draw[blue] (-0.25,-1.55) node{\scriptsize$b$};
      \draw[black] (-0.725,-1.075) node{\scriptsize$s$};
   \draw[black,thick,mid arrow] (0.77,-1.6)-- (2,-1.6);
   \draw[black,thick,mid arrow] (2,-2)--(0.77,-2);
   \draw[blue,thick] (0.58,-1.725)-- (1.9,-1.725);
   \draw[blue,thick] (1.9,-1.875)-- (0.58,-1.875);  

   \draw[black,thick] (1.98,-0.2)-- (1.98,-1.6);
    \draw[black,thick] (1.98,0.2)-- (1.98,0.6);
    \draw[black,thick] (1.98,-2)-- (1.98,-2.4);

    \draw[black,thick] (1.92,0.4)-- (2.04,0.4);
    \draw[black,thick] (1.92,-2.2)-- (2.04,-2.2);
      
   \draw[blue,thick] (1.9,-1.725) arc(90:-90:0.075 and 0.075);

   \filldraw[red] (0.8,0.2) circle (1.2pt) node {$ $};
    \filldraw[red] (0.8,-0.2) circle (1.2pt) node {$ $};

       \filldraw[red] (0.8,-1.6) circle (1.2pt) node {$ $};
      \filldraw[red] (0.8,-2) circle (1.2pt) node {$ $};
           \draw[brown] (1.3,-0.9) node{\X4};

   \filldraw (1,0.2) circle (0pt) node[above] {\scriptsize$X$};
   \filldraw (1,-2) circle (0pt) node[below] {\scriptsize$X$};
\end{tikzpicture}\times (-2)+
\begin{tikzpicture}[scale = 0.75,baseline={([yshift=-3.2pt]current bounding box.center)}]
   \draw[blue,thick] (-0.6,0) arc(180:8:0.6 and 0.6);
   \draw[blue,thick] (-0.6,0) arc(180:352:0.6 and 0.6);
   \draw[black,thick] (-0.8,0) arc(180:15:0.8 and 0.8);
   \draw[black,thick] (-0.8,0) arc(180:345:0.8 and 0.8);
      \draw[blue] (-0.25,0.25) node{\scriptsize$b$};
      \draw[black] (-0.725,0.725) node{\scriptsize$s$};
   \draw[black,thick,mid arrow] (0.77,0.2)-- (2,0.2);
   \draw[black,thick,mid arrow] (2,-0.2)--(0.77,-0.2);
   \draw[blue,thick] (0.58,0.075)-- (1.9,0.075);
   \draw[blue,thick] (1.9,-0.075)-- (0.58,-0.075);  
      
   \draw[blue,thick] (1.9,0.075) arc(90:-90:0.075 and 0.075);

      \draw[blue,thick] (-0.6,-1.8) arc(180:8:0.6 and 0.6);
   \draw[blue,thick] (-0.6,-1.8) arc(180:352:0.6 and 0.6);
   \draw[black,thick] (-0.8,-1.8) arc(180:15:0.8 and 0.8);
   \draw[black,thick] (-0.8,-1.8) arc(180:345:0.8 and 0.8);
      \draw[blue] (-0.25,-1.55) node{\scriptsize$b$};
      \draw[black] (-0.725,-1.075) node{\scriptsize$s$};
   \draw[black,thick,mid arrow] (0.77,-1.6)-- (2,-1.6);
   \draw[black,thick,mid arrow] (2,-2)--(0.77,-2);
   \draw[blue,thick] (0.58,-1.725)-- (1.9,-1.725);
   \draw[blue,thick] (1.9,-1.875)-- (0.58,-1.875);  

   \draw[black,thick] (1.98,-0.2)-- (1.98,-1.6);
    \draw[black,thick] (1.98,0.2)-- (1.98,0.6);
    \draw[black,thick] (1.98,-2)-- (1.98,-2.4);

    \draw[black,thick] (1.92,0.4)-- (2.04,0.4);
    \draw[black,thick] (1.92,-2.2)-- (2.04,-2.2);
      
   \draw[blue,thick] (1.9,-1.725) arc(90:-90:0.075 and 0.075);

   \filldraw[red] (0.8,0.2) circle (1.2pt) node {$ $};
    \filldraw[red] (0.8,-0.2) circle (1.2pt) node {$ $};

       \filldraw[red] (0.8,-1.6) circle (1.2pt) node {$ $};
      \filldraw[red] (0.8,-2) circle (1.2pt) node {$ $};
           \draw[brown] (1.3,-0.9) node{\X5};

   \filldraw (1,0.2) circle (0pt) node[above] {\scriptsize$X$};
   \filldraw (1,-1.6) circle (0pt) node[above] {\scriptsize$X$};
\end{tikzpicture}\ +\ \begin{tikzpicture}[scale = 0.75,baseline={([yshift=-3.2pt]current bounding box.center)}]
   \draw[blue,thick] (-0.6,0) arc(180:8:0.6 and 0.6);
   \draw[blue,thick] (-0.6,0) arc(180:352:0.6 and 0.6);
   \draw[black,thick] (-0.8,0) arc(180:15:0.8 and 0.8);
   \draw[black,thick] (-0.8,0) arc(180:345:0.8 and 0.8);
      \draw[blue] (-0.25,0.25) node{\scriptsize$b$};
      \draw[black] (-0.725,0.725) node{\scriptsize$s$};
   \draw[black,thick,mid arrow] (0.77,0.2)-- (2,0.2);
   \draw[black,thick,mid arrow] (2,-0.2)--(0.77,-0.2);
   \draw[blue,thick] (0.58,0.075)-- (1.9,0.075);
   \draw[blue,thick] (1.9,-0.075)-- (0.58,-0.075);  
      
   \draw[blue,thick] (1.9,0.075) arc(90:-90:0.075 and 0.075);

      \draw[blue,thick] (-0.6,-1.8) arc(180:8:0.6 and 0.6);
   \draw[blue,thick] (-0.6,-1.8) arc(180:352:0.6 and 0.6);
   \draw[black,thick] (-0.8,-1.8) arc(180:15:0.8 and 0.8);
   \draw[black,thick] (-0.8,-1.8) arc(180:345:0.8 and 0.8);
      \draw[blue] (-0.25,-1.55) node{\scriptsize$b$};
      \draw[black] (-0.725,-1.075) node{\scriptsize$s$};
   \draw[black,thick,mid arrow] (0.77,-1.6)-- (2,-1.6);
   \draw[black,thick,mid arrow] (2,-2)--(0.77,-2);
   \draw[blue,thick] (0.58,-1.725)-- (1.9,-1.725);
   \draw[blue,thick] (1.9,-1.875)-- (0.58,-1.875);  

   \draw[black,thick] (1.98,-0.2)-- (1.98,-1.6);
    \draw[black,thick] (1.98,0.2)-- (1.98,0.6);
    \draw[black,thick] (1.98,-2)-- (1.98,-2.4);

    \draw[black,thick] (1.92,0.4)-- (2.04,0.4);
    \draw[black,thick] (1.92,-2.2)-- (2.04,-2.2);
      
   \draw[blue,thick] (1.9,-1.725) arc(90:-90:0.075 and 0.075);

   \filldraw[red] (0.8,0.2) circle (1.2pt) node {$ $};
    \filldraw[red] (0.8,-0.2) circle (1.2pt) node {$ $};

       \filldraw[red] (0.8,-1.6) circle (1.2pt) node {$ $};
      \filldraw[red] (0.8,-2) circle (1.2pt) node {$ $};
           \draw[brown] (1.3,-0.9) node{\X6};

   \filldraw (1,-0.2) circle (0pt) node[below] {\scriptsize$X$};
   \filldraw (1,-2) circle (0pt) node[below] {\scriptsize$X$};
\end{tikzpicture}\ .
\end{aligned}
\end{equation}
As a result, the perturbation Page curve is a superposition of two-point functions on the entropy contour \eqref{eqn:entropycontour}. Two stages of the entropy then follows from the two regimes for two-point functions on the entropy contour (see numerical results in \cite{chen2020replica}).

\begin{figure}[t]
\centering
\includegraphics[width=0.5\linewidth]{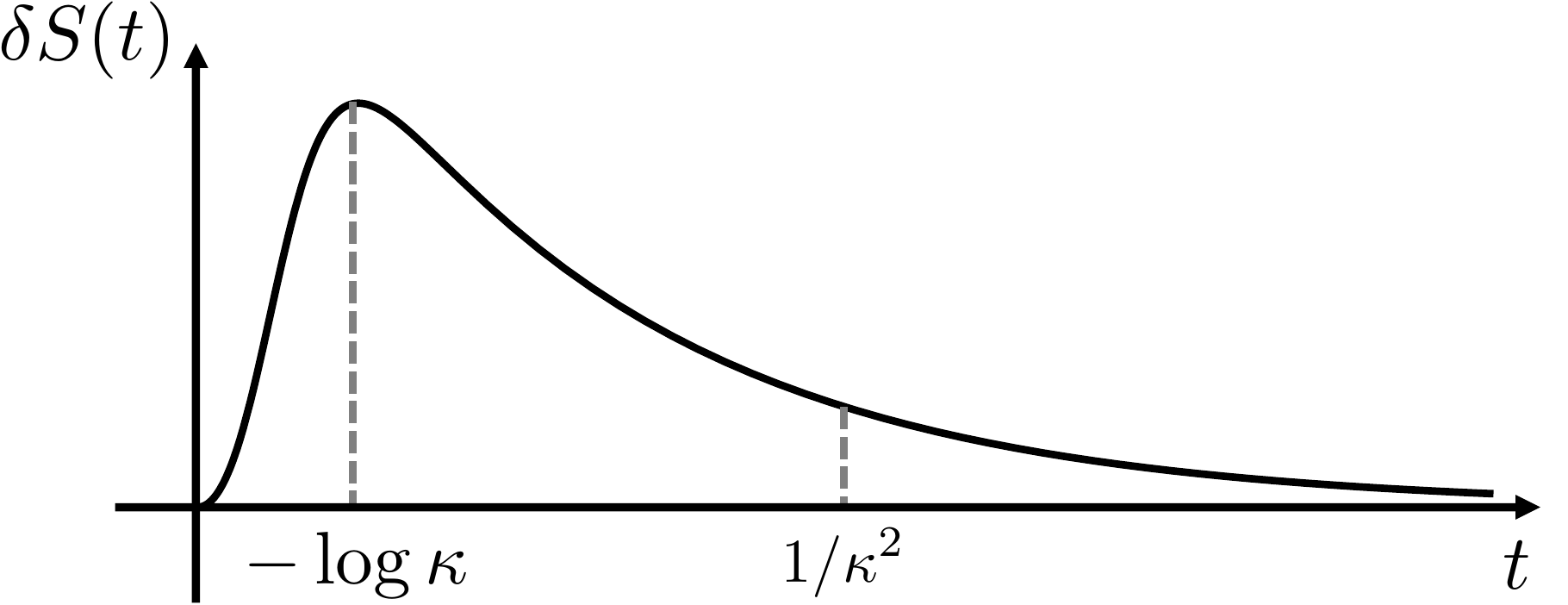}
\caption{A sketch of the perturbative Page curve as a function of evolution time $t$ for small system bath coupling $\kappa$ and arbitrary $n$. Here we choose some typical energy scales of system $s$ as the unit. For $t\ll -\log \kappa$, the curve is dominated by the early-time OTOC \cite{dadras2021perturbative}, which leads to an exponential growth $\sim \kappa^2 e^{\varkappa t}$ with quantum Lyapunov exponent $\varkappa$. In the long-time limit, it is governed by the energy relaxation $\sim e^{-\kappa^2 t}$. }\label{fig1}
\end{figure}

In the early-time regime, the system bath coupling can be taken into account perturbatively. We first consider the limit with $\kappa =0$, which gives:
\begin{equation}\label{eqn:earlyexp}
\text{\X1}=\text{\X2}=\text{\X4}=\langle X^2\rangle_{2\beta},\ \ \ \ \ \ \text{\X3}=\text{\X5}=\text{\X6}=\langle X(-i\beta)X(0)\rangle_{2\beta}. 
\end{equation} 
This leads to the cancellation between contributions $(\text{\X1},\text{\X2},\text{\X4})$ and $(\text{\X3},\text{\X5},\text{\X6})$. Similar relations hold for $t=0$ at finite $\kappa$, leading to $\delta S^{(2)}(0)=0$. On the other hand, for $t \rightarrow \infty$, we instead expect  
\begin{equation}\label{eqn:long_equivalent}
\begin{tikzpicture}[scale = 0.75,baseline={([yshift=-3.2pt]current bounding box.center)}]
   \draw[blue,thick] (-0.6,0) arc(180:8:0.6 and 0.6);
   \draw[blue,thick] (-0.6,0) arc(180:352:0.6 and 0.6);
   \draw[black,thick] (-0.8,0) arc(180:15:0.8 and 0.8);
   \draw[black,thick] (-0.8,0) arc(180:345:0.8 and 0.8);
      \draw[blue] (-0.25,0.25) node{\scriptsize$b$};
      \draw[black] (-0.725,0.725) node{\scriptsize$s$};
   \draw[black,thick,mid arrow] (0.77,0.2)-- (3,0.2);
   \draw[black,thick,mid arrow] (3,-0.2)--(0.77,-0.2);
   \draw[blue,thick] (0.58,0.075)-- (2.9,0.075);
   \draw[blue,thick] (2.9,-0.075)-- (0.58,-0.075);  
      
   \draw[blue,thick] (2.9,0.075) arc(90:-90:0.075 and 0.075);

      \draw[blue,thick] (-0.6,-1.8) arc(180:8:0.6 and 0.6);
   \draw[blue,thick] (-0.6,-1.8) arc(180:352:0.6 and 0.6);
   \draw[black,thick] (-0.8,-1.8) arc(180:15:0.8 and 0.8);
   \draw[black,thick] (-0.8,-1.8) arc(180:345:0.8 and 0.8);
      \draw[blue] (-0.25,-1.55) node{\scriptsize$b$};
      \draw[black] (-0.725,-1.075) node{\scriptsize$s$};
   \draw[black,thick,mid arrow] (0.77,-1.6)-- (3,-1.6);
   \draw[black,thick,mid arrow] (3,-2)--(0.77,-2);
   \draw[blue,thick] (0.58,-1.725)-- (2.9,-1.725);
   \draw[blue,thick] (2.9,-1.875)-- (0.58,-1.875);  

   \draw[black,thick] (2.98,-0.2)-- (2.98,-1.6);
    \draw[black,thick] (2.98,0.2)-- (2.98,0.6);
    \draw[black,thick] (2.98,-2)-- (2.98,-2.4);

    \draw[black,thick] (2.92,0.4)-- (3.04,0.4);
    \draw[black,thick] (2.92,-2.2)-- (3.04,-2.2);
      
   \draw[blue,thick] (2.9,-1.725) arc(90:-90:0.075 and 0.075);

   \filldraw[red] (0.8,0.2) circle (1.2pt) node {$ $};
    \filldraw[red] (0.8,-0.2) circle (1.2pt) node {$ $};

       \filldraw[red] (0.8,-1.6) circle (1.2pt) node {$ $};
      \filldraw[red] (0.8,-2) circle (1.2pt) node {$ $};

\end{tikzpicture}\ \ \ \approx \ \ \ 
\begin{tikzpicture}[scale = 0.75,baseline={([yshift=-3.2pt]current bounding box.center)}]
   \draw[blue,thick] (-0.6,0) arc(180:8:0.6 and 0.6);
   \draw[blue,thick] (-0.6,0) arc(180:352:0.6 and 0.6);
   \draw[black,thick] (-0.8,0) arc(180:15:0.8 and 0.8);
   \draw[black,thick] (-0.8,0) arc(180:345:0.8 and 0.8);
      \draw[blue] (-0.25,0.25) node{\scriptsize$b$};
      \draw[black] (-0.725,0.725) node{\scriptsize$s$};
   \draw[black,thick,mid arrow] (0.77,0.2)-- (3,0.2);
   \draw[black,thick,mid arrow] (3,-0.2)--(0.77,-0.2);
   \draw[blue,thick] (0.58,0.075)-- (2.9,0.075);
   \draw[blue,thick] (2.9,-0.075)-- (0.58,-0.075);  
      
   \draw[blue,thick] (2.9,0.075) arc(90:-90:0.075 and 0.075);
      \draw[black,thick] (3,-0.2) arc(-90:90:0.2 and 0.2);

      \draw[blue,thick] (-0.6,-1.8) arc(180:8:0.6 and 0.6);
   \draw[blue,thick] (-0.6,-1.8) arc(180:352:0.6 and 0.6);
   \draw[black,thick] (-0.8,-1.8) arc(180:15:0.8 and 0.8);
   \draw[black,thick] (-0.8,-1.8) arc(180:345:0.8 and 0.8);
      \draw[blue] (-0.25,-1.55) node{\scriptsize$b$};
      \draw[black] (-0.725,-1.075) node{\scriptsize$s$};
   \draw[black,thick,mid arrow] (0.77,-1.6)-- (3,-1.6);
   \draw[black,thick,mid arrow] (3,-2)--(0.77,-2);
   \draw[blue,thick] (0.58,-1.725)-- (2.9,-1.725);
   \draw[blue,thick] (2.9,-1.875)-- (0.58,-1.875);

   \draw[blue,thick] (2.9,-1.725) arc(90:-90:0.075 and 0.075);
      \draw[black,thick] (3,-2) arc(-90:90:0.2 and 0.2);

   \filldraw[red] (0.8,0.2) circle (1.2pt) node {$ $};
    \filldraw[red] (0.8,-0.2) circle (1.2pt) node {$ $};

       \filldraw[red] (0.8,-1.6) circle (1.2pt) node {$ $};
      \filldraw[red] (0.8,-2) circle (1.2pt) node {$ $};

\end{tikzpicture}\ \ \ ,
\end{equation}
unless we are interested in two-point functions near the final time $t$. This is consistent with the numerical result in \cite{chen2020replica}. One can understand \eqref{eqn:long_equivalent} by viewing the contour \eqref{eqn:entropycontour} as an evolution from right to the left: Initially, the system is described by some initial state specified by correlation functions near time $t$. The system is then evolved backward to time $0$, with a coupling to bath. Physically, we expect the coupling to bath thermalizes the system, which leads to the equality \eqref{eqn:long_equivalent}. Such a picture will be refined in later sections to derive a semi-classical Boltzmann for two-point functions on the entropy contour, which is valid for weakly interacting systems. \eqref{eqn:long_equivalent} suggests that at $t\rightarrow \infty$ we have
\begin{equation}\label{eqn:verylateexp}
\text{\X1}=\text{\X2}=\text{\X3}=\langle X^2\rangle_{\beta},\ \ \ \ \ \ \text{\X4}=\text{\X5}=\text{\X6}=0. 
\end{equation} 
This leads to $\delta S^{(2)}(\infty)=0$, consistent with the previous expectation.

The perturbative Page curve appears when two-point functions evolve from \eqref{eqn:earlyexp} to \eqref{eqn:verylateexp}. The decay of $\text{\X4},\text{\X5},\text{\X6}$ is a consequence of quantum many-body chaos. As an example, to the $\kappa^2$ order, $\text{\X4}$ receives the contribution from
\begin{equation}\label{eqn:pertb_Chaos}
\begin{tikzpicture}[scale = 0.75,baseline={([yshift=-3.2pt]current bounding box.center)}]
   \draw[blue,thick] (-0.6,0) arc(180:8:0.6 and 0.6);
   \draw[blue,thick] (-0.6,0) arc(180:352:0.6 and 0.6);
   \draw[black,thick] (-0.8,0) arc(180:15:0.8 and 0.8);
   \draw[black,thick] (-0.8,0) arc(180:345:0.8 and 0.8);
      \draw[blue] (-0.25,0.25) node{\scriptsize$b$};
      \draw[black] (-0.725,0.725) node{\scriptsize$s$};
   \draw[black,thick] (0.77,0.2)-- (2,0.2);
   \draw[black,thick] (2,-0.2)--(0.77,-0.2);
   \draw[blue,thick] (0.58,0.075)-- (1.9,0.075);
   \draw[blue,thick] (1.9,-0.075)-- (0.58,-0.075);  
      
   \draw[blue,thick] (1.9,0.075) arc(90:-90:0.075 and 0.075);

      \draw[blue,thick] (-0.6,-1.8) arc(180:8:0.6 and 0.6);
   \draw[blue,thick] (-0.6,-1.8) arc(180:352:0.6 and 0.6);
   \draw[black,thick] (-0.8,-1.8) arc(180:15:0.8 and 0.8);
   \draw[black,thick] (-0.8,-1.8) arc(180:345:0.8 and 0.8);
      \draw[blue] (-0.25,-1.55) node{\scriptsize$b$};
      \draw[black] (-0.725,-1.075) node{\scriptsize$s$};
   \draw[black,thick] (0.77,-1.6)-- (2,-1.6);
   \draw[black,thick] (2,-2)--(0.77,-2);
   \draw[blue,thick] (0.58,-1.725)-- (1.9,-1.725);
   \draw[blue,thick] (1.9,-1.875)-- (0.58,-1.875);  

   \draw[black,thick] (1.98,-0.2)-- (1.98,-1.6);
    \draw[black,thick] (1.98,0.2)-- (1.98,0.6);
    \draw[black,thick] (1.98,-2)-- (1.98,-2.4);

    \draw[black,thick] (1.92,0.4)-- (2.04,0.4);
    \draw[black,thick] (1.92,-2.2)-- (2.04,-2.2);
      
   \draw[blue,thick] (1.9,-1.725) arc(90:-90:0.075 and 0.075);

   \filldraw[red] (0.8,0.2) circle (1.2pt) node {$ $};

      \filldraw[red] (0.8,-2) circle (1.2pt) node {$ $};
           \draw[brown] (1.3,-0.9) node{\X4};

   \filldraw (1,0.2) circle (0pt) node[above] {\scriptsize$X$};
   \filldraw (1,-2) circle (0pt) node[below] {\scriptsize$X$};

  \draw[black,dashed] (1.6,0.2)-- (1.6,0.075);
  \filldraw[red] (1.6,0.2) circle (1.2pt) node {$ $};
    \filldraw[blue] (1.6,0.075) circle (1.2pt) node {$ $};

      \draw[black,dashed] (1.8,-0.2)-- (1.8,-0.075);
  \filldraw[red] (1.8,-0.2) circle (1.2pt) node {$ $};
    \filldraw[blue] (1.8,-0.075) circle (1.2pt) node {$ $};

\end{tikzpicture}\ \ \propto \ \ \kappa^2\sum_{ij}\int_{12}^t~\langle O_b^i(t_1-i\beta)O_b^j(t_2) \rangle_\beta \langle O^i_s(t_1-2i\beta)X(-2i\beta)O^j_s(t_2-i\beta)X(0)\rangle_{2\beta},
\end{equation}
which contains OTOC of system $s$. We thus expect: 
\begin{equation}
\text{\X4}\sim \langle X^2\rangle_{2\beta}\times(1 -\# \kappa^2 e^{\varkappa t}),\ \ \ \ \ \ 
\text{\X5}\sim \text{\X6}\sim\langle X(-i\beta)X(0)\rangle_{2\beta} \times(1 -\# \kappa^2 e^{\varkappa t})
\end{equation}
 As a result, we have $\delta S^{(2)}\propto \kappa^2 e^{\varkappa t}$ for $\varkappa t \ll -\log \kappa$. Similar contributions have been calculated using toy models in \cite{dadras2021perturbative}. When $\kappa^2 e^{\varkappa t}\sim O(1)$, we should sum up terms with multiple scramblons as in \cite{gu2022two,sizenewpaper,Zhang:2022knu}, which leads to the decay of $\text{\X4},\text{\X5},\text{\X6}$.

 On the other hand, the evolution of $\text{\X1},\text{\X2},\text{\X3}$ is dominated by the thermalization process. We expect the energy relaxation has the smallest relaxation rate $\lambda_0\sim \kappa^2$. As a result, we expect that
 \begin{equation}\label{eqn:lateexp}
 \begin{aligned}
&\text{\X1}=\text{\X2}\approx\langle X^2\rangle_{\beta}+\big(\langle X^2\rangle_{2\beta}-\langle X^2\rangle_{\beta}\big)e^{-\lambda_0 t},\\
&\text{\X3}\approx\langle X^2\rangle_{\beta}+\big(\langle X(-i\beta)X(0)\rangle_{2\beta}-\langle X^2\rangle_{\beta}\big)e^{-\lambda_0 t}.
\end{aligned}
 \end{equation}
 This leads to the exponential decay of $\delta S^{(2)}(t)$ in the second stage. In the following sections of this paper, we will try to give a more quantitative description of above discussions using generalized Boltzmann equations for systems with quasi-particles, or perturbative calculations for 0+1-d large-$N$ systems. 

\section{Boltzmann Equation for R\'enyi Entropies}\label{secIII}
For systems with quasi-particles, the thermalization process is usually studied by semi-classical Boltzmann equations. In particular, hydrodynamical parameters, such as diffusion constant, heat conductivity, and viscosity, can be computed under the Boltzmann equation approximation \cite{pitaevskii2012physical}. In quantum systems, Boltzmann equations can be derived using the Schwinger-Dyson equation on the Keldysh contour with the assumption that the quantum distribution function is slow varying \cite{kamenev2011field}. Later, it is realized that OTO-correlations can also be studied using generalized Boltzmann equations on doubled Keldysh contour \cite{ALEINER2016378}, which is a non-linear version of the kinetic equation at weak coupling \cite{Stanford:2015owe,zhang2021obstacle}. 

In this section, motivated by the backward evolution picture, we derive a set of new Boltzmann equations on the entropy contour \eqref{eqn:entropycontour}. These equations naturally capture both quantum many-body chaos and quantum thermalization, which gives a complete description of the perturbative Page curve for R\'enyi entropies in systems with quasi-particles. We will mainly focus on $n=2$. Generalizations to $n\geq 3$ are straightforward, while we find no simple approach to take the limit of $n\rightarrow 1$.

\subsection{The derivation}

Our derivation below is generally valid for systems at weak coupling, with arbitrary flavors of particles and minor differences between bosons and fermions. Nevertheless, a concrete model can be helpful. We consider a $D$-dimensional fermionic system with an SYK-like interaction
\begin{equation}
H_s=\sum_{k,i} \epsilon_s(k)c_s^i(k)^\dagger c_s^i(k)+\frac{1}{4V}\sum_{ijkl}\sum_{\{k_a\}}J_{ijkl}c_s^i(k_4)^\dagger c_s^j(k_3)^\dagger c_s^k(k_2) c_s^l(k_1).
\end{equation} 
Here $i,j,k,l=1,2,...,N$, $k_4=-k_1-k_2-k_3$ due to the momentum conservation, and random interaction strength $J_{ijkl}$ are independent random Gaussian variables with $\overline{|J_{ijkl}|^2}=2J^2/N^3$. Similarly, we model the bath $b$ as free fermions with SYK-like couplings to the system $s$
\begin{equation}\label{eqn:boltz_Hint}
H_b+H_{sb}=\sum_{k,m} \epsilon_b(k)c_b^m(k)^\dagger c_b^m(k)+\frac{1}{2V}\sum_{impq}\sum_{\{k_a\}}\Big[\kappa_{impq}c_s^i(k_4)^\dagger c_b^m(k_3)^\dagger c_b^p(k_2) c_b^q(k_1)+\text{H.C.}\Big].
\end{equation}
Here $m,p,q=1,2,...,M$, and the random system-bath interactions satisfy $\overline{|\kappa_{imnp}|^2}=2\kappa^2/M^3$. We assume $M\gg N$ and the bath correlation functions receive no correction from finite system-bath coupling $\kappa$. We choose the initial perturbation as $c^1_s(k)+c^1_s(k)^\dagger$, and \eqref{eqn:sixcontributions} becomes a summation over fermion two-point functions.

We begin with a useful identity that shows the causality of the R\'enyi entropy contour, implying the possibility of a differential equation description. Diagrammatically, the identity states that 
\begin{equation}\label{eqn:contourBoltz}
\begin{tikzpicture}[scale = 0.75,baseline={([yshift=-3.2pt]current bounding box.center)}]
   \draw[blue,thick] (-0.6,0) arc(180:8:0.6 and 0.6);
   \draw[blue,thick] (-0.6,0) arc(180:352:0.6 and 0.6);
   \draw[black,thick] (-0.8,0) arc(180:15:0.8 and 0.8);
   \draw[black,thick] (-0.8,0) arc(180:345:0.8 and 0.8);
      \draw[blue] (-0.25,0.25) node{\scriptsize$b$};
      \draw[black] (-0.725,0.725) node{\scriptsize$s$};
   \draw[black,thick,far arrow] (0.77,0.2)-- (3,0.2);
   \draw[black,thick,near arrow] (3,-0.2)--(0.77,-0.2);
   \draw[blue,thick] (0.58,0.075)-- (2.9,0.075);
   \draw[blue,thick] (2.9,-0.075)-- (0.58,-0.075);  
      
   \draw[blue,thick] (2.9,0.075) arc(90:-90:0.075 and 0.075);

      \draw[blue,thick] (-0.6,-1.8) arc(180:8:0.6 and 0.6);
   \draw[blue,thick] (-0.6,-1.8) arc(180:352:0.6 and 0.6);
   \draw[black,thick] (-0.8,-1.8) arc(180:15:0.8 and 0.8);
   \draw[black,thick] (-0.8,-1.8) arc(180:345:0.8 and 0.8);
      \draw[blue] (-0.25,-1.55) node{\scriptsize$b$};
      \draw[black] (-0.725,-1.075) node{\scriptsize$s$};
   \draw[black,thick,far arrow] (0.77,-1.6)-- (3,-1.6);
   \draw[black,thick,near arrow] (3,-2)--(0.77,-2);
   \draw[blue,thick] (0.58,-1.725)-- (2.9,-1.725);
   \draw[blue,thick] (2.9,-1.875)-- (0.58,-1.875);  

   \draw[black,thick] (2.98,-0.2)-- (2.98,-1.6);
    \draw[black,thick] (2.98,0.2)-- (2.98,0.6);
    \draw[black,thick] (2.98,-2)-- (2.98,-2.4);

    \draw[black,thick] (2.92,0.4)-- (3.04,0.4);
    \draw[black,thick] (2.92,-2.2)-- (3.04,-2.2);
      
   \draw[blue,thick] (2.9,-1.725) arc(90:-90:0.075 and 0.075);

   \filldraw[red] (1.5,0.2) circle (1.2pt) node {$ $};
    \filldraw[red] (1.5,-0.2) circle (1.2pt) node {$ $};

       \filldraw[red] (1.5,-1.6) circle (1.2pt) node {$ $};
      \filldraw[red] (1.5,-2) circle (1.2pt) node {$ $};

  \draw[brown] (1.9,-0.5) node{\scriptsize$d_2$};
  \draw[brown] (1.9,0.5) node{\scriptsize$u_2$};

  \draw[brown] (1.9,-2.3) node{\scriptsize$d_1$};
  \draw[brown] (1.9,-1.3) node{\scriptsize$u_1$};

    \draw[black] (0.93,0.5) node{\scriptsize$-\infty$};
    \draw[black] (2.8,0.5) node{\scriptsize$0$};
\end{tikzpicture}\ \ \ =\ \ \ 
\begin{tikzpicture}[scale = 0.75,baseline={([yshift=-3.2pt]current bounding box.center)}]
   \draw[blue,thick] (-0.6,0) arc(180:8:0.6 and 0.6);
   \draw[blue,thick] (-0.6,0) arc(180:352:0.6 and 0.6);
   \draw[black,thick] (-0.8,0) arc(180:15:0.8 and 0.8);
   \draw[black,thick] (-0.8,0) arc(180:345:0.8 and 0.8);
      \draw[blue] (-0.25,0.25) node{\scriptsize$b$};
      \draw[black] (-0.725,0.725) node{\scriptsize$s$};
   \draw[black,thick,mid arrow] (0.77,0.2)-- (2,0.2);
   \draw[black,thick,mid arrow] (2,-0.2)--(0.77,-0.2);
   \draw[blue,thick] (0.58,0.075)-- (1.9,0.075);
   \draw[blue,thick] (1.9,-0.075)-- (0.58,-0.075);  
      
   \draw[blue,thick] (1.9,0.075) arc(90:-90:0.075 and 0.075);

      \draw[blue,thick] (-0.6,-1.8) arc(180:8:0.6 and 0.6);
   \draw[blue,thick] (-0.6,-1.8) arc(180:352:0.6 and 0.6);
   \draw[black,thick] (-0.8,-1.8) arc(180:15:0.8 and 0.8);
   \draw[black,thick] (-0.8,-1.8) arc(180:345:0.8 and 0.8);
      \draw[blue] (-0.25,-1.55) node{\scriptsize$b$};
      \draw[black] (-0.725,-1.075) node{\scriptsize$s$};
   \draw[black,thick,mid arrow] (0.77,-1.6)-- (2,-1.6);
   \draw[black,thick,mid arrow] (2,-2)--(0.77,-2);
   \draw[blue,thick] (0.58,-1.725)-- (1.9,-1.725);
   \draw[blue,thick] (1.9,-1.875)-- (0.58,-1.875);  

   \draw[black,thick] (1.98,-0.2)-- (1.98,-1.6);
    \draw[black,thick] (1.98,0.2)-- (1.98,0.6);
    \draw[black,thick] (1.98,-2)-- (1.98,-2.4);

    \draw[black,thick] (1.92,0.4)-- (2.04,0.4);
    \draw[black,thick] (1.92,-2.2)-- (2.04,-2.2);
      
   \draw[blue,thick] (1.9,-1.725) arc(90:-90:0.075 and 0.075);

   \filldraw[red] (0.8,0.2) circle (1.2pt) node {$ $};
    \filldraw[red] (0.8,-0.2) circle (1.2pt) node {$ $};

       \filldraw[red] (0.8,-1.6) circle (1.2pt) node {$ $};
      \filldraw[red] (0.8,-2) circle (1.2pt) node {$ $};

\end{tikzpicture}.
\end{equation}
This is because the evolution commutes with the initial thermal density matrix, and is an analog of the cancellation between forward and backward evolutions on the Keldysh contour. Consequently, we only need to study two-point functions on a contour with a long evolution time and focus on branches with real-time evolutions. We label these branches by forward/backward index $\eta=u/d$ and replica index $n=1,2$, which is combined into $\alpha=\eta_n$ for conciseness. This is illustrated in \eqref{eqn:contourBoltz}. Motivated by the ordering of the contour, we take the convention to represent fields as a vector with the ordering $(d_1,u_1,d_2,u_2)$.

The Green's function $G_{\alpha\beta}(k;t_1,t_2)=\langle \psi^i_{s,\alpha}(k,t_1)\bar{\psi}^i_{s,\beta}(k,t_2)\rangle$ is introduced with fermionic fields $\psi^i_{\eta,\alpha}$ of the system $s$. For spatially homogeneous initial conditions, the Schwinger-Dyson equation in the momentum space reads 
\begin{equation}\label{eqn:SD1}
\begin{aligned}
(-1)^\alpha(\partial_{t_1}+i\epsilon_s(k))G_{\alpha \beta}-\Sigma_{\alpha\gamma}\circ G_{\gamma \beta}&=I_{\alpha\beta},\\
(-1)^\beta(-\partial_{t_2}+i\epsilon_s(k))G_{\alpha \beta}-G_{\alpha\gamma}\circ \Sigma_{\gamma \beta}&=I_{\alpha\beta}.
\end{aligned}
\end{equation}
Here we have introduced $(-1)^\alpha=1$ for $u_n$ branches and $(-1)^\alpha=-1$ for $d_n$ branches. The self-energy is given by melon diagrams 
\begin{equation}\label{eqn:SD2}
\begin{aligned}
\Sigma_{\alpha \beta}(k,t_1,t_2)=-\mathcal{P}_{\alpha\beta}\int \frac{dk_1^D}{(2\pi)^D}\frac{dk_2^D}{(2\pi)^D}\Big[&J^2G_{\alpha\beta}(k_2;t_1,t_2)G_{\alpha\beta}(k_3;t_1,t_2)G_{\beta\alpha}(k_1;t_2,t_1)\\
&+\kappa^2 \tilde G_{\alpha\beta}(k_2;t_1,t_2)\tilde G_{\alpha\beta}(k_3;t_1,t_2)\tilde G_{\beta\alpha}(k_1;t_2,t_1)\Big].
\end{aligned}
\end{equation}
Here $k_3=k_1+k-k_2$ by momentum conservation. $\mathcal{P}_{\alpha\beta}=-1$ if both $\alpha$ and $\beta$ are $u$ or $d$ contours. Otherwise, we have $\mathcal{P}_{\alpha\beta}=1$. $\tilde G_{\alpha\beta}(k;t_1,t_2)$ is the bath Green's function, introduced similarly to $G_{\alpha\beta}(k;t_1,t_2)$, and is diagonal in the replica space.

The Boltzmann equation is an approximation of the Schwinger-Dyson equation \cite{kamenev2011field}. The first step is to parametrize Green's functions $G_{\alpha \beta}$ by introducing quantum distribution functions. We use the motivation from the early-time \eqref{eqn:earlyexp} and the late-time limit \eqref{eqn:verylateexp}. In the early-time regime, we can show that
\begin{equation}\label{eqn:Boltzearly}
G_{\alpha\beta}=e^{-i\epsilon_s(k) t_{12}}
\begin{pmatrix}
\theta(t_{21})-n_{2\beta}(k)&-w_{2\beta}(k)&-w_{2\beta}(k)&1-n_{2\beta}(k)\\
w_{2\beta}(k)&\theta(t_{12})-n_{2\beta}(k)&n_{2\beta}(k)&-w_{2\beta}(k)\\
w_{2\beta}(k)&1-n_{2\beta}(k)&\theta(t_{21})-n_{2\beta}(k)&-w_{2\beta}(k)\\
n_{2\beta}(k)&w_{2\beta}(k)&w_{2\beta}(k)&\theta(t_{12})-n_{2\beta}(k)
\end{pmatrix}.
\end{equation}
Here $n_{2\beta}(k)=[{e^{2\beta (\epsilon_s(k)-\mu)}+1}]^{-1}$ is the Fermi-Dirac distribution with inverse temperature $2\beta$ and $w_{2\beta}=[2\cosh (\beta(\epsilon_s(k)-\mu))]^{-1}$. On the other hand, in the late-time limit, the relation \eqref{eqn:long_equivalent} gives
\begin{equation}\label{eqn:late-quasiparticle}
G_{\alpha\beta}=e^{-i\epsilon_s(k) t_{12}}
\begin{pmatrix}
\theta(t_{21})-n_{\beta}(k)&-1+n_{\beta}(k)&0&0\\
n_{\beta}(k)&\theta(t_{12})-n_{\beta}(k)&0&0\\
0&0&\theta(t_{21})-n_{\beta}(k)&-1+n_{\beta}(k)\\
0&0&n_{\beta}(k)&\theta(t_{12})-n_{\beta}(k)
\end{pmatrix}.
\end{equation}
Furthermore, the bath Green's function $\tilde G_{\alpha\beta}$ can be obtained by replacing $\epsilon_s(k)$ with $\epsilon_b(k)$ in \eqref{eqn:late-quasiparticle}. Comparing Green's functions in two limits, we find the evolution of Green's functions can be viewed as the evolution of $n_{2\beta}(k)$ and $w_{2\beta}(k)$. After reducing the number of variables using the symmetry between branches, we parametrize the Green's function at finite center-of-mass time $T=\frac{t_1+t_2}{2}$ as 
\begin{equation}\label{eqn:parametrization}
G_{\alpha\beta}=e^{-i\epsilon_s(k) t_{12}}
\begin{pmatrix}
\theta(t_{21})-f_0(k,T)&-f_2(k,T)&-f_3(k,T)&f_5(k,T)\\
f_1(k,T)&\theta(t_{12})-f_0(k,T)&f_4(k,T)&-f_3(k,T)\\
f_3(k,T)&f_5(k,T)&\theta(t_{21})-f_0(k,T)&-f_2(k,T)\\
f_4(k,T)&f_3(k,T)&f_1(k,T)&\theta(t_{12})-f_0(k,T)
\end{pmatrix}.
\end{equation}
Compared to distribution functions on the (doubled) Keldysh contour, there are three different distribution functions even if we focus on a single contour, which reflects the non-trivial twist operator on the system $s$. 

To derive the Boltzmann equation for $f_z(k,t)$ $(z=0,1,...,5)$, we start by taking the superposition of \eqref{eqn:SD1}, which gives
\begin{equation}
\partial_TG_{\alpha\beta}(k,t_1,t_2)=\int_3~(-1)^\alpha\Sigma_{\alpha\gamma}(k,t_1,t_3)G_{\gamma\beta}(k,t_3,t_2)-(-1)^\beta G_{\alpha\gamma}(k,t_1,t_3)\Sigma_{\gamma\beta}(k,t_3,t_2).
\end{equation}
We further set $t_1=t_2$. When $f_z$ are slow varying, we make approximations to the integral over $t_3$ by completing the integral with $f_z(k,\frac{t_1+t_3}{2})\approx f_z(k,\frac{t_1+t_2}{2})$. This leads to 
\begin{equation}
\begin{aligned}
\partial_tG_{\alpha\beta}(k,t,t)=\int_{-\infty}^\infty dt_r~&\Bigg[(-1)^\alpha\Sigma_{\alpha\gamma}\left(k,t+\frac{t_r}{2},t-\frac{t_r}{2}\right)G_{\gamma\beta}\left(k,t-\frac{t_r}{2},t+\frac{t_r}{2}\right)\\&-(-1)^\beta G_{\alpha\gamma}\left(k,t+\frac{t_r}{2},t-\frac{t_r}{2}\right)\Sigma_{\gamma\beta}\left(k,t-\frac{t_r}{2},t+\frac{t_r}{2}\right)\Bigg].
\end{aligned}
\end{equation}
The L.H.S. is just the time derivative of distribution functions $f_z$. Now using \eqref{eqn:SD2} and \eqref{eqn:parametrization}, it is straightforward to show that the integral over time $t$ only imposes the energy conservation of the scattering process\footnote{One may worry about the additional $\theta(t)$ function appearing in the diagonal elements of $G_{\alpha \beta}$. Fortunately, their contributions of $G_{u_nu_n}$ and $G_{d_nd_n}$ always cancel out.}, which appears in Boltzmann equations \cite{kamenev2011field,ALEINER2016378}. After tedious by straightforward calculations, we find the result read
\begin{equation}\label{eqn:Boltzmannequation}
\partial_t f_z(k,t)=J^2\int d\mathcal{M}_s~\text{St}_z[f_{z'}(k,t)]+\kappa^2\int d\mathcal{M}_b~\tilde{\text{St}}_z[f_{z'}(k,t)].
\end{equation}
Here $d\mathcal{M}_{\xi}$ is the phase space integral with momentum and energy conservation
\begin{equation}
d\mathcal{M}_{\xi}\equiv \prod_{i=1}^3\left[\frac{dk_i^D}{(2\pi)^D}\right]~(2\pi)^{D+1}\delta^{(D)}(k+k_1-k_2-k_3)\delta(\epsilon_{s}(k)+\epsilon_{\xi}(k_1)-\epsilon_{\xi}(k_2)-\epsilon_{\xi}(k_3)).
\end{equation}
$\text{St}_z[f]$ are combinations of distribution functions
\begin{equation}
\begin{aligned}
\text{St}_0[f]=&{f_1}({k_2}) {f_1}({k_3}) {f_2}(k) {f_2}({k_1})-{f_1}(k) {f_1}({k_1}) {f_2}({k_2})
   {f_2}({k_3})\\&+{f_4}({k_2}) {f_4}({k_3}) {f_5}(k) {f_5}({k_1})-{f_4}(k) {f_4}({k_1})
   {f_5}({k_2}) {f_5}({k_3}),\\
\text{St}_1[f]=&{f_1}(k) ({f_0}({k_1}) ({f_0}({k_2}) (1-2
   {f_0}({k_3}))+{f_0}({k_3})-1)+{f_0}({k_2}) {f_0}({k_3}))\\&+(1-2 {f_0}(k)) {f_1}({k_2}) {f_1}({k_3}) {f_2}({k_1})-2 {f_3}(k) {f_4}({k_2}) {f_4}({k_3})
   {f_5}({k_1})\\&+2 {f_3}({k_1}) {f_3}({k_2}) {f_3}({k_3}) {f_4}(k),\\
\text{St}_2[f]=&  {f_2}(k) (-({f_0}({k_1}) ({f_0}({k_2}) (1-2
   {f_0}({k_3}))+{f_0}({k_3})-1)+{f_0}({k_2}) {f_0}({k_3})))\\&+(2 {f_0}(k)-1) {f_1}({k_1}) {f_2}({k_2}) {f_2}({k_3})-2 {f_3}(k) {f_4}({k_1}) {f_5}({k_2})
   {f_5}({k_3})\\&+2 {f_3}({k_1}) {f_3}({k_2}) {f_3}({k_3}) {f_5}(k),\\
\text{St}_3[f]=&   {f_1}({k_1}) {f_2}({k_2}) {f_2}({k_3}) {f_4}(k)+{f_1}({k_2}) {f_1}({k_3}) {f_2}({k_1})
   {f_5}(k)\\&-{f_1}(k) {f_4}({k_1}) {f_5}({k_2}) {f_5}({k_3})-{f_2}(k) {f_4}({k_2}) {f_4}({k_3})
   {f_5}({k_1}),\\
\text{St}_4[f]=&{f_4}(k) ({f_0}({k_1}) (-2 {f_0}({k_2})
   {f_0}({k_3})+{f_0}({k_2})+{f_0}({k_3})-1)+{f_0}({k_2}) {f_0}({k_3}))\\&+(1-2 {f_0}(k)) {f_4}({k_2}) {f_4}({k_3}) {f_5}({k_1})+2 {f_1}({k_2})
   {f_1}({k_3}) {f_2}({k_1}) {f_3}(k)\\&-2 {f_1}(k) {f_3}({k_1}) {f_3}({k_2}) {f_3}({k_3}),\\
\text{St}_5[f]=&{f_5}(k) ({f_0}({k_1}) ((2 {f_0}({k_2})-1)
   {f_0}({k_3})-{f_0}({k_2})+1)-{f_0}({k_2}) {f_0}({k_3}))\\&+(2 {f_0}(k)-1) {f_4}({k_1}) {f_5}({k_2}) {f_5}({k_3})+2 {f_1}({k_1}) {f_2}({k_2})
   {f_2}({k_3}) {f_3}(k)\\&-2 {f_2}(k) {f_3}({k_1}) {f_3}({k_2}) {f_3}({k_3}).
   \end{aligned}
\end{equation}
Here we keep the $t$ dependence implicit for conciseness. $\tilde{\text{St}}_z[f]$ can be obtained by replacing $f_3(k_i)=f_4(k_i)=f_5(k_i)=0$ and $f_0(k_i)=f_1(k_i)=1-f_2(k_i)=n_\beta(k_i)$ with dispersion $\epsilon_b(k)$. This is the main result of this section. The equation \eqref{eqn:Boltzmannequation} should be evolved backwardly in time, with initial conditions determined by comparing \eqref{eqn:Boltzearly} and \eqref{eqn:parametrization}. By using \eqref{eqn:sixcontributions} and \eqref{eqn:contourBoltz}, the perturbation of the second R\'enyi entropy is given by
\begin{equation}
\delta S^{(2)}(t)=2-2(f_1(k,-t)+f_2(k,-t))-2f_4(k,-t)-2f_5(k,-t)+4f_3(k,-t).
\end{equation}
Here we relabel the time $t$ such that the twist operator is inserted at $t=0$. This is illustrated in \eqref{eqn:long_equivalent}. 

Here are some comments:
\begin{itemize}
  \item It is straightforward to check that when $f_3=f_4=f_5=0$ and $f_0=f_1=1-f_2=f(k,t)$, these set of equations are reduced to the traditional Boltzmann equation, with
\begin{equation}
\text{St}_0=\text{St}_1=-\text{St}_2=f(k_2)f(k_3)(1-f(k_1))(1-f(k))-(k,k_1)\leftrightarrow (k_2,k_3).
\end{equation}
In particular, $f(k)=n_\beta(k)$ is the steady-state solution. This is consistent with the expectation of the late-time solution \eqref{eqn:late-quasiparticle}. 

  \item Without the coupling to the bath ($\kappa=0$), the contour \eqref{eqn:contourBoltz} is equivalent to doubled Keldysh contour \cite{ALEINER2016378}. As a result, the initial value \eqref{eqn:Boltzearly} is a unstable steady state at $\kappa=0$. A finite $\kappa$ serves as an initial perturbation that induces the OTO-correlations. This is consistent with the perturbative analysis \eqref{eqn:pertb_Chaos}.

  \item The contribution of $\tilde{\text{St}}_z[f]$ is linear in $f_z[k,t]$. This is a consequence of the choice of our system-bath coupling \eqref{eqn:boltz_Hint}, which contains a single operator in system $s$. More general couplings lead to a non-linear system-bath scattering $\tilde{\text{St}}_z[f]$.

\end{itemize}

\begin{figure}[t]
\centering
\includegraphics[width=1.\linewidth]{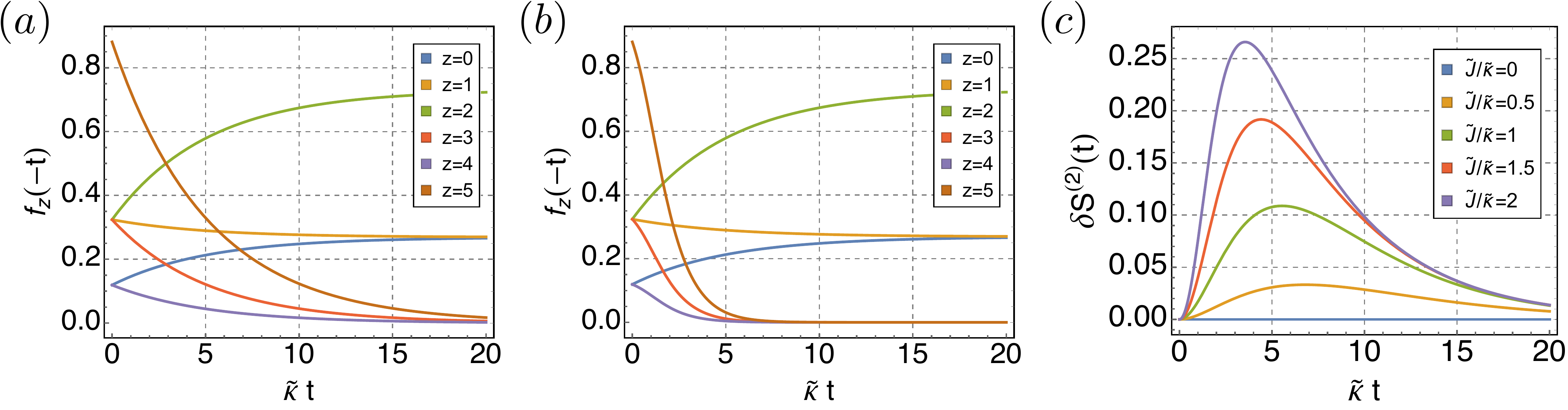}
\caption{Results for the perturbative Page curve in the $D=0$ example with $\beta \epsilon=1$: (a). The analytical solution of the Boltzmann equation $f_z(t)$ for $\tilde{J}/\tilde{\varkappa}=0$; (b). The numerical solution of the Boltzmann equation $f_z(t)$ for $\tilde{J}/\tilde{\varkappa}=2$; (c). The entropy $\delta S^{(2)}(t)$ as a function of time $t$ for different $\tilde{J}/\tilde{\kappa}$. The result shows the separation of time scale for $\tilde{\kappa}\ll \tilde{J}$ as in Figure \ref{fig1}. }\label{fig2}
\end{figure}

\subsection{Example: non-local fermionic systems}
In general, such non-linear integro-differential equations should be studied numerically to determine the perturbative Page curve. In this subsection, we focus on the simplest $D=0$ case as an example, where all momentum indices drop out. We further choose $\epsilon_s=\epsilon_b=\epsilon$ to ensure the energy conservation in the scattering process. An concrete example is the SYK model at high temperature. As observed in \cite{Qi:2020ian}, the energy conservation contains a divergence for $D=0$, which should be regularized by the quasi-particle lifetime $\Gamma^{-1}$. Without diving into the details, here we simply introduce 
\begin{equation}
\tilde{J}^2\equiv J^2 \int d\mathcal{M}_s \sim J^2/\Gamma,\ \ \ \ \ \ \tilde{\kappa}^2\equiv\kappa^2 \int d\mathcal{M}_b \sim \kappa^2/\Gamma.
\end{equation} 

We first consider the $J=0$ case. With our system-bath coupling \eqref{eqn:boltz_Hint}, it is known that the system $s$ is not many-body chaotic \cite{Zhang:2019fcy}. As a result, the evolution of $f_z(t)$ is only driven by relaxation due to the bath $b$. Thanks to the linearity of the Boltzmann equation, analytical solutions can be obtained in the closed-form:
\begin{equation}
\begin{aligned}
&f_0(t)=\frac{-(w-1) e^{\lambda_0 t}+w+w^{-1}}{(w+1) \left(w+w^{-1}\right)},\ \ \ \ \ \ f_1(t)=\frac{(1-w^{-1}) e^{\lambda_0 t}+w+w^{-1}}{(w+1) \left(w+w^{-1}\right)},\\
&f_2(t)=wf_0(t),\ \ \ \ \ \ \ \ \ \ \ f_3(t)=wf_4(t)=w^{-1}f_5(t)=\frac{e^{\lambda_0 t}}{w+w^{-1}}.
\end{aligned}
\end{equation}
Here we have defined $w=e^{\beta \epsilon}$ and $\lambda_0=\tilde{\kappa}^2/(w+w^{-1}+2)$. The result shows that $f_z(t)$ decay exponentially for negative $t$, to the thermalized value \eqref{eqn:late-quasiparticle}. However, direct calculation shows that we have $\delta S^{(2)}(t)=0$. This is consistent with the fact that the initial increase of $\delta S^{(2)}(t)$ is driven by many-body chaos, which is absent for $J=0$.

Now we consider finite $\tilde{J}/\tilde{\kappa}$ by performing a numerical study of the Boltzmann equation \eqref{eqn:Boltzmannequation}. For $\tilde{J}\gg \tilde{\kappa}$, $f_3(t),f_4(t),f_5(t)$, which are off-diagonal in the replica space, now decays much more rapid comparing to the relaxation due to the coupling to bath $b$, as shown in Figure \ref{fig2} (b). As a result, the entropy shows a rapid increase, followed by slow relaxation, consistent with the sketch in Figure \ref{fig1}.  

\section{Results for Large-N Models in 0+1-d}\label{secIV}
In the previous section, we analyzed $\delta S^{(n)}(t)$ by deriving Boltzmann equations for systems with quasi-particles. In this section, we instead consider general systems with all-to-all interactions, which may describe non-Fermi liquids without quasi-particles \cite{Chowdhury:2021qpy}. A concrete example of such system $s$ is the celebrated SYK model at low temperature \cite{sachdev1993gapless,kitaevsimple,maldacena2016remarks}\footnote{The entropy in SYK-like models has also been studied in \cite{Chen:2020atj,Gu:2017njx,Liu:2017kfa,zhang2020entanglement,Zhang:2020kia,Su:2020quk,Haldar:2020ymg,Chen:2019qqe,jian2021phase,Dadras:2019tcz}. }. We also keep the choice of operators $X$, $O_s^i$ general. In the early-time regime, we derive general results for arbitrary $n$ including both the exponential growth and the saturation of $\delta S^{(n)}(t)$. However, the analytical continuation to $n=1$ turns out to be hard. In the late-time regime, we present results for both R\'enyi entropies and the Von Neumann entropy. 

\subsection{Early-time scramblon effective theory}
We firstly consider the early-time regime which is dominated by the decay of inter-replica two-point functions $\text{\X4},\text{\X5},\text{\X6}$ due to many-body chaos. For sufficiently small $\kappa$, we can safely assume all intr-replica Green's functions are the same as $\kappa=0$, and only interactions induced by exchanging scramblons are important, which allows us to compute inter-replica two-point functions using the scramblon effective theory developed in \cite{gu2022two,sizenewpaper}. For simplicity, we will
perform the calculation for a bosonic system, but the final result also works for fermionic systems.

We first consider integrating out the bath on the entropy contour with $n$ replicas. This leads to a pictorial representation:
\begin{equation}\exp\left[-(n-1)S^{(n)}(t,0)\right]\ \ \ =\ \ \ 
\begin{tikzpicture}[scale=0.5,baseline={([yshift=-3pt]current bounding box.center)}]

\draw[thick,black] (-60pt,5pt)--(-20pt,5pt);
\draw[thick,black] (-20pt,-5pt)--(-60pt,-5pt);

\draw[thick,black] (20pt,5pt)--(60pt,5pt);
\draw[thick,black] (60pt,-5pt)--(20pt,-5pt);

\draw[thick,black] (5pt,-60pt)--(5pt,-20pt);
\draw[thick,black] (-5pt,-20pt)--(-5pt,-60pt);

\draw[thick,black] (5pt,20pt)--(5pt,60pt);
\draw[thick,black] (-5pt,60pt)--(-5pt,20pt);

\draw[thick,black] (20pt,5pt) arc(90:270:5pt and 5pt);
\draw[thick,black] (5pt,20pt) arc(0:-180:5pt and 5pt);

\draw[thick,black] (-20pt,5pt) arc(90:-90:5pt and 5pt);
\draw[thick,black] (5pt,-20pt) arc(0:180:5pt and 5pt);

\draw[thick,black] (60pt,5pt) arc(0:90:55pt and 55pt);
\draw[thick,black] (60pt,-5pt) arc(0:-90:55pt and 55pt);
\draw[thick,black] (-60pt,5pt) arc(180:90:55pt and 55pt);
\draw[thick,black] (-60pt,-5pt) arc(-180:-90:55pt and 55pt);

\draw[black] (0,-70pt) node{\scriptsize$0$};
\draw[black] (-70pt,0) node{\scriptsize$\beta$};
\draw[black] (0,70pt) node{\scriptsize$2\beta$};
\draw[black] (70pt,0) node{\scriptsize$3\beta$};

\draw[black] (0,0pt) node{\scriptsize$t$};

\draw[thick,black,dashed] (45pt,5pt) arc(0:90:40pt and 40pt);
\draw[thick,black,dashed] (45pt,-5pt) arc(0:-90:40pt and 40pt);
\draw[thick,black,dashed] (-45pt,5pt) arc(180:90:40pt and 40pt);
\draw[thick,black,dashed] (-45pt,-5pt) arc(-180:-90:40pt and 40pt);

\draw[thick,black,dashed] (30pt,5pt) arc(0:90:25pt and 25pt);
\draw[thick,black,dashed] (30pt,-5pt) arc(0:-90:25pt and 25pt);
\draw[thick,black,dashed] (-30pt,5pt) arc(180:90:25pt and 25pt);
\draw[thick,black,dashed] (-30pt,-5pt) arc(-180:-90:25pt and 25pt);

\end{tikzpicture}\ \ \ .
\end{equation}
Here we take $n=4$ as an example. The Lorentzian time is directed towards the center of the diagram. The dashed lines represent the interaction induced by integrating out bath $b$ (we neglect terms that are not relevant to OTO-correlations as in \cite{dadras2021perturbative}), which reads 
\begin{equation}
S_{\text{int}}=\kappa^2\sum_{i=1}^N\sum_{k=0}^{n-1}\int dt_1 dt_2~O^i_s(t_1-i(k+1)\beta^-)O^i_s(t_2-ik\beta^+)\tilde{G}_{ud}(t_{12}).
\end{equation}
Here $\beta^\pm=\beta\pm 0^+$. We begin with the computation of \X4, which contains two $X$ operators in different replicas. Generalizing to general $n$, we find
\begin{equation}\label{eqn:perturbation4}
\text{\X4}\ \ \ =\ \ \ 
\begin{tikzpicture}[scale=0.55,baseline={([yshift=-3pt]current bounding box.center)}]

\draw[thick,black] (-60pt,5pt)--(-20pt,5pt);
\draw[thick,black] (-20pt,-5pt)--(-60pt,-5pt);

\draw[thick,black] (20pt,5pt)--(60pt,5pt);
\draw[thick,black] (60pt,-5pt)--(20pt,-5pt);

\draw[thick,black] (5pt,-60pt)--(5pt,-20pt);
\draw[thick,black] (-5pt,-20pt)--(-5pt,-60pt);

\draw[thick,black] (5pt,20pt)--(5pt,60pt);
\draw[thick,black] (-5pt,60pt)--(-5pt,20pt);

\draw[thick,black] (20pt,5pt) arc(90:270:5pt and 5pt);
\draw[thick,black] (5pt,20pt) arc(0:-180:5pt and 5pt);

\draw[thick,black] (-20pt,5pt) arc(90:-90:5pt and 5pt);
\draw[thick,black] (5pt,-20pt) arc(0:180:5pt and 5pt);

\draw[thick,black] (60pt,5pt) arc(0:90:55pt and 55pt);
\draw[thick,black] (60pt,-5pt) arc(0:-90:55pt and 55pt);
\draw[thick,black] (-60pt,5pt) arc(180:90:55pt and 55pt);
\draw[thick,black] (-60pt,-5pt) arc(-180:-90:55pt and 55pt);

\draw[thick,black,dashed] (-45pt,5pt) arc(180:90:40pt and 40pt);
\draw[thick,black,dashed] (-45pt,-5pt) arc(-180:-90:40pt and 40pt);

\draw[thick,black,dashed] (-30pt,5pt) arc(180:90:25pt and 25pt);
\draw[thick,black,dashed] (-30pt,-5pt) arc(-180:-90:25pt and 25pt);

\filldraw[red] (-5pt,-60pt) circle (3pt) node {$ $};
\filldraw[red] (-5pt,60pt) circle (3pt) node {$ $};

\end{tikzpicture}
\ \ \ +\ \ \ 
\begin{tikzpicture}[scale=0.55,baseline={([yshift=-3pt]current bounding box.center)}]

\draw[thick,black] (-60pt,5pt)--(-20pt,5pt);
\draw[thick,black] (-20pt,-5pt)--(-60pt,-5pt);

\draw[thick,black] (20pt,5pt)--(60pt,5pt);
\draw[thick,black] (60pt,-5pt)--(20pt,-5pt);

\draw[thick,black] (5pt,-60pt)--(5pt,-20pt);
\draw[thick,black] (-5pt,-20pt)--(-5pt,-60pt);

\draw[thick,black] (5pt,20pt)--(5pt,60pt);
\draw[thick,black] (-5pt,60pt)--(-5pt,20pt);

\draw[thick,black] (20pt,5pt) arc(90:270:5pt and 5pt);
\draw[thick,black] (5pt,20pt) arc(0:-180:5pt and 5pt);

\draw[thick,black] (-20pt,5pt) arc(90:-90:5pt and 5pt);
\draw[thick,black] (5pt,-20pt) arc(0:180:5pt and 5pt);

\draw[thick,black] (60pt,5pt) arc(0:90:55pt and 55pt);
\draw[thick,black] (60pt,-5pt) arc(0:-90:55pt and 55pt);
\draw[thick,black] (-60pt,5pt) arc(180:90:55pt and 55pt);
\draw[thick,black] (-60pt,-5pt) arc(-180:-90:55pt and 55pt);

\draw[thick,black,dashed] (45pt,5pt) arc(0:90:40pt and 40pt);
\draw[thick,black,dashed] (-45pt,-5pt) arc(-180:-90:40pt and 40pt);

\draw[thick,black,dashed] (30pt,5pt) arc(0:90:25pt and 25pt);
\draw[thick,black,dashed] (-30pt,-5pt) arc(-180:-90:25pt and 25pt);

\filldraw[red] (-5pt,-60pt) circle (3pt) node {$ $};
\filldraw[red] (60pt,5pt) circle (3pt) node {$ $};

\end{tikzpicture}\ \ \ +\ \ \ 
\begin{tikzpicture}[scale=0.55,baseline={([yshift=-3pt]current bounding box.center)}]

\draw[thick,black] (-60pt,5pt)--(-20pt,5pt);
\draw[thick,black] (-20pt,-5pt)--(-60pt,-5pt);

\draw[thick,black] (20pt,5pt)--(60pt,5pt);
\draw[thick,black] (60pt,-5pt)--(20pt,-5pt);

\draw[thick,black] (5pt,-60pt)--(5pt,-20pt);
\draw[thick,black] (-5pt,-20pt)--(-5pt,-60pt);

\draw[thick,black] (5pt,20pt)--(5pt,60pt);
\draw[thick,black] (-5pt,60pt)--(-5pt,20pt);

\draw[thick,black] (20pt,5pt) arc(90:270:5pt and 5pt);
\draw[thick,black] (5pt,20pt) arc(0:-180:5pt and 5pt);

\draw[thick,black] (-20pt,5pt) arc(90:-90:5pt and 5pt);
\draw[thick,black] (5pt,-20pt) arc(0:180:5pt and 5pt);

\draw[thick,black] (60pt,5pt) arc(0:90:55pt and 55pt);
\draw[thick,black] (60pt,-5pt) arc(0:-90:55pt and 55pt);
\draw[thick,black] (-60pt,5pt) arc(180:90:55pt and 55pt);
\draw[thick,black] (-60pt,-5pt) arc(-180:-90:55pt and 55pt);

\draw[thick,black,dashed] (45pt,-5pt) arc(0:-90:40pt and 40pt);

\draw[thick,black,dashed] (-45pt,-5pt) arc(-180:-90:40pt and 40pt);

\draw[thick,black,dashed] (30pt,-5pt) arc(0:-90:25pt and 25pt);

\draw[thick,black,dashed] (-30pt,-5pt) arc(-180:-90:25pt and 25pt);

\filldraw[red] (-5pt,-60pt) circle (3pt) node {$ $};
\filldraw[red] (5pt,-60pt) circle (3pt) node {$ $};

\end{tikzpicture}\ \ \ .
\end{equation}
The corresponding contribution to entropy reads $\delta S^{(n)}_{\text{\X4}}=-\frac{n}{n-1}\text{\X4}$. \eqref{eqn:perturbation4} can be computed using the scramblon effective theory \cite{gu2022two}: The initial insertion of two $X$ operators in the past create a perturbation propagating forwardly in time, with a distribution of the perturbation strength $y$ described by $h_{X}^\text{A}(y,-i\tau)_{n\beta}$. Such a perturbation is observed by pairs of $O_s$ operators in the future, which changes their two-point functions from $G_{O_s}(t_{12})_{n\beta}$ to $f_{O_s}^\text{R}\left(C^{-1}e^{i\delta }e^{\varkappa \frac{t_1+t_2}{2}}y,t_{12}\right)_{n\beta}$. Here $C\propto N$ is the coefficient of OTOC and $\delta$ is fixed by the imaginary-time configuration \cite{kitaev2018soft,gu2019relation}\footnote{To be precise, here $\varkappa=\varkappa(n\beta)$ and $C=C(n\beta)$ are defined at inverse temperature $n\beta$. }. This leads to the result
\begin{equation}
\begin{aligned}
\text{\X4}&=\sum_{k=1}^{n-1}\int_0^\infty dy~h_{X}^\text{A}(y,-i(k+1)\beta)_{n\beta}\exp(-S_{\text{\X4}}^k),\\
\frac{S_{\text{\X4}}^k}{\kappa^2N}&=\int_{12}\tilde{G}_{ud}(t_{12})\left[2G_{O_s}(t_{12}-i\beta)_{n\beta}-f_{O_s}^\text{R}\left(\frac{e^{i\delta_4 }e^{\varkappa \frac{t_1+t_2}{2}}y}{C},t_{12}-i\beta\right)_{n\beta}-(\delta_4\rightarrow -\delta_4)\right].
\end{aligned}
\end{equation}
Here we have introduced $\delta_4=\frac{\varkappa \beta}{2}(\frac{n}{2}-k)$. To take the limit of $N \rightarrow \infty$, we expand $f^{\text{R/A}}(z,t)=G(t)-z\Upsilon^\text{R/A}(t)+O(z^2)$ \cite{gu2022two}. The result reads
\begin{equation}
\frac{S_{\text{\X4}}^k}{\kappa^2}=y\frac{2N\cos\delta_4(k)}{C}\int_0^t dt_1\int_0^t dt_2~\tilde{G}_{ud}(t_{12})e^{\varkappa\frac{t_1+t_2}{2}}\Upsilon^\text{R}_{O_s}(t_{12}-i\beta)_{n\beta}\equiv y \cos\delta_4(k) \Lambda(t).
\end{equation}
Here we can estimate $\Lambda(t)$ as
\begin{equation}
\begin{aligned}
\Lambda(t)&\approx \frac{2N}{C} \int_{-\infty}^\infty dt_{12} \int_0^tdT~e^{\varkappa T}\tilde{G}_{ud}(t_{12})\Upsilon^\text{R}_{O_s}(t_{12}-i\beta)_{n\beta}\\
&=\frac{2N}{\varkappa C} (e^{\varkappa t}-1) \int_{-\infty}^\infty dt_{12}~\tilde{G}_{ud}(t_{12})\Upsilon^\text{R}_{O_s}(t_{12}-i\beta)_{n\beta},
\end{aligned}
\end{equation}
which grows exponentially in time with exponent $\varkappa$. The integral over $y$ becomes a Laplace transform of $h_X^\text{A}$. Using the definition that $f^\text{R/A}(z,t)=\int_0^\infty dy~h^\text{R/A}(y,t) e^{-zy}$ \cite{gu2022two}, we find 
\begin{equation}\label{eqn:perturbationres4}
\begin{aligned}
\text{\X4}&=\sum_{k=1}^{n-1}f_{X}^\text{A}(\kappa^2\cos\delta_4(k) \Lambda(t),-i(k+1)\beta)_{n\beta}.
\end{aligned}
\end{equation}

The generalizations of \X5 and \X6 can be computed similarly, which contributes to the entropy as $\delta S^{(n)}_{\text{\X5}}=\frac{n}{2(n-1)}\text{\X5}$ and $\delta S^{(n)}_{\text{\X6}}=\frac{n}{2(n-1)}\text{\X6}$. Here we only present the result:
\begin{equation}
\begin{aligned}
\text{\X5}&=\sum_{k=1}^{n-1}f_{X}^\text{A}(\kappa^2\cos\delta_4(k)e^{i\varkappa \beta/2} \Lambda(t),-ik\beta)_{n\beta},\\
\text{\X6}&=\sum_{k=1}^{n-1}f_{X}^\text{A}(\kappa^2\cos\delta_4(k)e^{-i\varkappa \beta/2} \Lambda(t),-ik\beta)_{n\beta}.
\end{aligned}
\end{equation}
Summing up contributions \X1, \X2, \X3, we arrive at the early-time result of $\delta S^{(n)}$:
\begin{equation}
\begin{aligned}\label{eqn:earlyperturbres}
\delta S^{(n)}&=\frac{n}{n-1}\Big(G_X(0)_{n\beta}-G_X(-i\beta)_{n\beta}-\sum_{k=1}^{n-1}f_{X}^\text{A}(\kappa^2\cos\delta_4(k) \Lambda(t),-i(k+1)\beta)_{n\beta}\\
&+\frac{1}{2}\sum_{k=1}^{n-1}\big(f_{X}^\text{A}(\kappa^2\cos\delta_4(k)e^{\frac{i\varkappa \beta}{2}} \Lambda(t),-ik\beta)_{n\beta}+f_{X}^\text{A}(\kappa^2\cos\delta_4(k)e^{-\frac{i\varkappa \beta}{2}} \Lambda(t),-ik\beta)_{n\beta}\big)\Big).
\end{aligned}
\end{equation}
This is the main result of this subsection. As an example, for $n=2$ we have
\begin{equation}
\begin{aligned}\label{eqn:earlyperturbresn2}
\delta S^{(2)}=&2\Big(G_X(0)_{2\beta}-G_X(-i\beta)_{2\beta}-f_{X}^\text{A}(\kappa^2 \Lambda(t),0)_{2\beta}\\&+\frac{f_{X}^\text{A}(\kappa^2 \Lambda(t)e^{i\varkappa \beta/2},-i\beta)_{2\beta}
+f_{X}^\text{A}(\kappa^2 \Lambda(t)e^{-i\varkappa \beta/2},-i\beta)_{2\beta}}{2}\Big).
\end{aligned}
\end{equation}
For the SYK model in the large-$q$ limit and the (complex) Brownian SYK model, $f^\text{A}_\chi$ has been computed in closed-form \cite{gu2022two,complexnewpaper}. At early time $\varkappa^2 \Lambda(t) \ll 1$, The entropy \eqref{eqn:earlyperturbres} grow exponentially, consistent with our general analysis in \eqref{eqn:pertb_Chaos}. The result saturates to $\frac{n}{n-1}(G_X(0)_{n\beta}-G_X(-i\beta)_{n\beta})$ for $\kappa^2 \Lambda(t) \gtrsim 1$, after which the entropy dynamics is dominated by the relaxation process, as analyzed in the next subsection. It is straightforward to the $n\rightarrow 1$ limit for the saturation value:
\begin{equation}
\delta S_\text{sat}=\left.\frac{G_X(0)_{n\beta}-G_X(-i\beta)_{n\beta}}{n-1}\right|_{n\rightarrow 1}=\beta \langle [X,H_s]X\rangle_\beta=\beta~\text{tr} [H_s (e^{-i\epsilon X} \rho e^{i\epsilon X}-\rho)]+o(\epsilon^2).
\end{equation}
As a result, $\delta S_\text{sat}$ is equal to the inverse temperature times the energy increase due to the external impulse. This is just the thermodynamical relation $\delta S=\beta \delta E$. As a result, $\delta S_\text{sat}$ matches the coarse-grained entropy.

\subsection{Late-time perturbation theory}
Now we consider the late-time regime where inter-replica correlations are zero. Consequently, different replicas decouple from each other and we can focus on a single replica:
\begin{equation}
\begin{tikzpicture}[scale = 0.75,baseline={([yshift=-3.2pt]current bounding box.center)}]
   \draw[blue,thick] (-0.6,0) arc(180:8:0.6 and 0.6);
   \draw[blue,thick] (-0.6,0) arc(180:352:0.6 and 0.6);
   \draw[black,thick] (-0.8,0) arc(180:15:0.8 and 0.8);
   \draw[black,thick] (-0.8,0) arc(180:345:0.8 and 0.8);
   \draw[black,thick,mid arrow] (0.77,0.2)-- (3,0.2);
   \draw[black,thick,mid arrow] (3,-0.2)--(0.77,-0.2);
   \draw[blue,thick] (0.58,0.075)-- (2.9,0.075);
   \draw[blue,thick] (2.9,-0.075)-- (0.58,-0.075);  
      
   \draw[blue,thick] (2.9,0.075) arc(90:-90:0.075 and 0.075);
   \draw[black,thick] (2.9,0.075) arc(90:-90:0.075 and 0.075);

   \filldraw[red] (0.8,0.2) circle (1.2pt) node {$ $};
    \filldraw[red] (0.8,-0.2) circle (1.2pt) node {$ $};

   \draw[black,thick,dotted] (3,0.2) arc(90:-90:0.2 and 0.2);

\end{tikzpicture}
\end{equation}
Nevertheless, we should keep in mind that the initial value $G_{ud}=\langle X(-i\beta)X(0)\rangle_{n\beta}$ due to the presence of other replicas. This is essential to see not the full late-time entropy matches the coarse-grained entropy:
\begin{equation}
\left.\partial_n \text{\X3}\right|_{n=1}=\left.\partial_n\langle X e^{iHt}e^{-(n-1)\beta H_s}e^{-iHt}X\rangle_\beta\right|_{n=1}=\beta \langle X H_s(t)X \rangle.
\end{equation}
Adding similar contributions from \X1 and \X2, we find $\delta S(t)=\beta \delta E(t)$. 

In the remaining part of this subsection, we will study the perturbation of $G_{ud}$ due to the system-bath coupling to the $\kappa^2$ order. The result validates \eqref{eqn:lateexp} and fixes the energy relaxation rate $\lambda_0$. To the order of $\kappa^2$, we have 
\begin{equation}\label{eqn:perturblate}
\delta \text{\X3}=\kappa^2\sum_i\sum_{\alpha\beta}\mathcal{P}_{\alpha
\beta}\int_0^t dt_1\int_0^t dt_2 ~\langle X_{u}(0)X_{d}(0)O^i_{s,\alpha}(t_1)O^i_{s,\beta}(t_2)\rangle _{c,n\beta}\langle O^i_{b,\alpha}(t_1)O^i_{b,\beta}(t_2)\rangle _{n\beta}
\end{equation}
Here $\alpha,\beta=u/d$. For different $(\alpha,\beta)$, the real-time connected four-point functions in \eqref{eqn:perturblate} can be expressed by the imaginary-time ordered four-point function $\mathcal{F}(t_1,t_2;t_3,t_4)_{n\beta}$ as
\begin{equation}
\begin{aligned}
(u,u)=&\theta(t_{12})\mathcal{F}(t_1-i\beta^+,t_2-i\beta;-i\beta^-,0)_{n\beta}+\theta(t_{21})\mathcal{F}(t_2-i\beta^+,t_1-i\beta;-i\beta^-,0)_{n\beta},\\
(d,d)=&\theta(t_{12})\mathcal{F}(t_2-in\beta,t_1-in\beta^-;-i\beta,0)_{n\beta}+\theta(t_{21})\mathcal{F}(t_1-in\beta,t_2-in\beta^-;-i\beta,0)_{n\beta},\\
(u,d)=&\mathcal{F}(t_2-in\beta,t_1-i\beta;-i\beta^-,0)_{n\beta},\ \ \ \ \ \ \ \ \ 
(d,u)=\mathcal{F}(t_1-in\beta,t_2-i\beta;-i\beta^-,0)_{n\beta}.
\end{aligned}
\end{equation}
Summing up all these contributions, the result reads
\begin{equation}\label{eqn:devlateper}
\begin{aligned}
\delta \text{\X3}=&2\kappa^2 \int_{t_2>t_1} \tilde{G}_{ud}(t_{12})\left(\mathcal{F}(t_2-in\beta,t_1-i\beta;-i\beta^-,0)_{n\beta}-\mathcal{F}(t_2-i\beta^+,t_1-i\beta;-i\beta^-,0)_{n\beta}\right)\\
&+2\kappa^2\int_{t_1>t_2} \tilde{G}_{ud}(t_{12})\left(\mathcal{F}(t_2-in\beta,t_1-i\beta;-i\beta^-,0)_{n\beta}-\mathcal{F}(t_2-in\beta^+,t_1-in\beta;-i\beta^-,0)_{n\beta}\right).
\end{aligned}
\end{equation}
This result vanishes in the limit of $n\rightarrow 1$, which leads to a finite contribution to the Von Neumann entropy. Expanding near $n=1$, we find
\begin{equation}\label{eqn:devlateper2}
\begin{aligned}
\frac{\delta \text{\X3}}{n-1}=&2\kappa^2(-i\beta) \int dt_1 dt_2 ~ \partial_{t_1}\tilde{G}_{ud}(t_{12})\mathcal{F}(t_2-i\beta^+,t_1-i\beta;-i\beta^-,0)_{\beta}.
\end{aligned}
\end{equation}

To proceed, we assume that the energy fluctuation contributes to the connected four-point function as
\begin{equation}\label{eqn:guessF}
\mathcal{F}(t_1,t_2;t_3,t_4)_\beta\approx C_\beta \partial_\beta G_{O_s}(t_{12})\partial_\beta G_{X}(t_{34}),\ \ \ \ \ \ \text{for} \ \ \ t_1,t_2\gg t_3, t_4.
\end{equation}
Physically, the insertion of $X$ operators creates an energy increase, which is probed by $O_s$ operators. It has been shown that \eqref{eqn:guessF} is valid in the conformal limit of the SYK model. This leads to 
\begin{equation}\label{eqn: late-time per res}
\begin{aligned}
\frac{\delta \text{\X3}}{n-1}&\approx 2\kappa^2(-i\beta) t\int_{-\infty}^{\infty} dt_{12}  ~ \partial_{t_1}\tilde{G}_{ud}(t_{12})C_\beta \partial_\beta G_{O_s}(t_{21})\left.\partial_\beta G_{X}(t_{34})\right|_{t_{34}->-i\beta}\\
&\equiv -\lambda_0 t \beta \left.\partial_\beta G_{X}(t_{34})\right|_{t_{34}->-i\beta}.
\end{aligned}
\end{equation}
Noticing that $$\langle X(-i\beta) X(0)\rangle_{n\beta}-\langle X^2 \rangle_\beta=-(n-1)\beta  \langle H X(-i\beta) X(0)\rangle_{\beta}=(n-1)\beta \left.\partial_\beta G_{X}(t_{34})\right|_{t_{34}->-i\beta},$$
the result \eqref{eqn: late-time per res} matches the expansion of \eqref{eqn:lateexp} for $\lambda_0 t \ll 1$. We thus identify
\begin{equation}
\lambda_0=2\kappa^2i \int_{-\infty}^{\infty} dt_{12}  ~ \partial_{t_1}\tilde{G}_{ud}(t_{12})C_\beta \partial_\beta G_{O_s}(t_{21}).
\end{equation}

\section{Discussions}\label{secFinal}
In this work, we study the perturbative Page curve in open quantum systems induced by an external impulse. The entropy dynamics exhibit a separation of time scales for small system-bath coupling, with an early-time exponential growth due to quantum many-body chaos, and a late-time relaxation due to the system-bath coupling. To provide a quantitative description, we first derive the generalized Boltzmann equation for the perturbation of R\'enyi entropies, which is valid for systems with quasi-particles. The effects of many-body chaos and energy relaxation are naturally encoded in distinct quantum distribution functions. To understand strongly correlated systems, we further study general large-$N$ systems in 0+1-d, where quasi-particles may be absent. In the early-time regime, we use the scramblon effective theory to derive the increase of R\'enyi entropies. In the late-time regime, we perform a perturbative analysis to determine the relaxation rate and demonstrate that the entropy tracks the coarse-grained entropy.

There are several potential extensions to the current work: Firstly the stability of the Boltzmann equation is of vital importance. It would be worthwhile to investigate whether some analog of $H$-theorems can be derived for generalized Boltzmann equations on the entropy contour. Another intriguing question is whether analogous Boltzmann equations exist for perturbations of Von Neumann entropies, which necessitates considering the limit as $n\rightarrow 1$. Exploring the evolution of generalized Boltzmann equations in systems where R\'enyi entropies can be experimentally measured, such as the Bose-Hubbard model \cite{Ho:2015woa}, could yield interesting insights. Finally, applying generalized Boltzmann equations to systems with repeated measurements may lead to a new perspective of the measurement-induced phase transitions.

\section*{Acknowledgment}
We thank Yu Chen, Pouria Dadras, Ruihua Fan, and Alexei Kitaev for helpful discussions.

\vspace{10pt}
\textit{Note Added.} When completing this work, we became aware of a related study on the entanglement dynamics, independently conducted by Yu Chen \cite{Yunewpaper}.

\bibliography{Page Curve.bbl}

\end{document}